\documentclass[fleqn,usenatbib]{mnras}

\usepackage[T1]{fontenc}

\DeclareRobustCommand{\VAN}[3]{#2}
\let\VANthebibliography\thebibliography
\def\thebibliography{\DeclareRobustCommand{\VAN}[3]{##3}\VANthebibliography}

\usepackage{graphicx}	
\usepackage{amsmath}	
\usepackage{amssymb}	
\usepackage{caption}
\usepackage{subcaption}
\usepackage{orcidlink}
\usepackage{newtxtext,newtxmath}

\title[AO 0235+164]{A Near Magnetic-to-kinetic Energy Equipartition Flare from the Relativistic Jet in AO 0235+164 during 2013-2019}

\author[Cheong et al.]{
Whee Yeon Cheong,$^{1,2}$$^{\orcidlink{0009-0002-1871-5824}}$
Sang-Sung Lee,$^{1,2}$\thanks{E-mail: sslee@kasi.re.kr}$^{\orcidlink{0000-0002-6269-594X}}$
Sang-Hyun Kim,$^{1,2}$$^{\orcidlink{0000-0001-7556-8504}}$
Sincheol Kang,$^{2}$$^{\orcidlink{0000-0002-0112-4836}}$
Jae Young Kim,$^{3,4}$$^{\orcidlink{0000-0001-8229-7183}}$
\newauthor
Bindu Rani,$^{5,2,6}$$^{\orcidlink{0000-0001-5711-084X}}$
Anthony C. S. Readhead,$^{7}$
Sebastian Kiehlmann,$^{8,9}$
Anne L\"ahteenm\"aki,$^{10,11}$
Merja Tornikoski,$^{10}$
\newauthor
Joni Tammi,$^{10}$
Venkatessh Ramakrishnan,$^{12,10}$
Iv\'an Agudo,$^{13}$
Antonio Fuentes,$^{13}$
Efthalia Traianou,$^{4,13}$
\newauthor
Juan Escudero,$^{13}$
Clemens Thum,$^{14}$
Ioannis Myserlis,$^{14}$$^{\orcidlink{0000-0003-3025-9497}}$
Carolina Casadio,$^{8,9}$
and Mark Gurwell$^{15}$$^{\orcidlink{0000-0003-0685-3621}}$
\\
$^{1}$University of Science and Technology, 217 Gajeong-ro, Yuseong-gu, Daejeon 34113, Republic of Korea\\
$^{2}$Korea Astronomy and Space Science Institute, 776 Daedeok-daero, Yuseong-gu, Daejeon 34055, Republic of Korea\\
$^{3}$Department of Astronomy and Atmospheric Sciences, Kyungpook National University, Daegu 702-701, Republic of Korea\\
$^{4}$Max-Planck-Institut f\"ur Radioastronomie, Auf dem H\"ugel 69, D-53121 Bonn, Germany\\
$^{5}$NASA Goddard Space Flight Center, Greenbelt, MD 20771, USA\\
$^{6}$Department of Physics, American University, Washington, DC 20016, USA\\
$^{7}$Owens Valley Radio Observatory, California Institute of Technology, Pasadena, CA 91125, USA\\
$^{8}$Institute of Astrophysics, Foundation for Research and Technology-Hellas, GR-71110 Heraklion, Greece\\
$^{9}$Department of Physics, Univ. of Crete, GR-70013 Heraklion, Greece\\
$^{10}$Aalto University Mets\"ahovi Radio Observatory, Mets\"ahovintie 114, 02540 Kylm\"al\"a, Finland\\
$^{11}$Aalto University Department of Electronics and Nanoengineering, P.O. BOX 15500, FI-00076 AALTO, Finland\\
$^{12}$Finnish Centre for Astronomy with ESO (FINCA), University of Turku, Vesilinnantie 5, 20014 University of Turku, Finland\\
$^{13}$Instituto de Astrof\'isica de Andaluc\'ia-CSIC, Glorieta de la Astronom\'ia, E-18008, Granada, Spain\\
$^{14}$Institut de Radioastronomie Millim\'{e}trique, Avenida Divina Pastora, 7, Local 20, E--18012 Granada, Spain\\
$^{15}$Center for Astrophysics \textbar Harvard \& Smithsonian, Cambridge, MA 02138 USA
}

\date{Accepted XXX. Received YYY; in original form ZZZ}

\pubyear{2023}

\begin{document}
\label{firstpage}
\pagerange{\pageref{firstpage}--\pageref{lastpage}}
\maketitle

\begin{abstract}
We present the multiwavelength flaring activity of the blazar AO 0235+164 during its recent active period from 2013 to 2019. From a discrete correlation function (DCF) analysis, we find a significant ($>95\%$) correlation between radio and $\gamma$-ray light curves with flares at longer wavelengths following flares at shorter wavelengths. We identify a new jet component in 43~GHz VLBA data that was ejected from the radio core on MJD~$57246^{+26}_{-30}$ (2015 August 12), during the peak of the 2015 radio flare. From the analysis of the jet component, we derived a Doppler factor of $\delta_{\rm var}=28.5\pm8.4$, a bulk Lorentz factor of $\Gamma=16.8^{+3.6}_{-3.1}$, and an intrinsic viewing angle of $\theta_{\rm v}=1.42^{+1.07}_{-0.52}\textrm{ degrees}$. Investigation of the quasi-simultaneous radio data revealed a partially absorbed spectrum with the turnover frequency varying in the range of $10-70$~GHz and the peak flux density varying in the range of $0.7-4\textrm{ Jy}$. We find the synchrotron self-absorption magnetic field strength to be $B_{\rm SSA}=15.3^{+12.6}_{-14.0}\textrm{ mG}$ at the peak of the 2015 radio flare, which is comparable to the equipartition magnetic field strength of $B_{\rm EQ}=43.6^{+10.6}_{-10.4}\textrm{ mG}$ calculated for the same epoch. Additional analysis of the radio emission region in the relativistic jet of AO 0235+164 suggests that it did not significantly deviate from equipartition during its recent flaring activity. 
\end{abstract}

\begin{keywords}
BL Lacertae objects: individual: AO 0235+164 -- galaxies: active -- galaxies: jets -- radio continuum: galaxies
\end{keywords}



\section{Introduction}

Powered by accretion onto supermassive black holes, active galactic nuclei (AGN) produce copious amounts of high-energy emission. Understanding the physical conditions and processes responsible for the production of high-energy particles and the emission of $\gamma$-rays is one of the most challenging quests of high-energy astrophysics. We present here a thorough analysis of the multi-frequency observation of a blazar, AO 0235+164, to probe the physical process(es) responsible for $\gamma$-ray emission in the source. \\
\indent{}AO 0235+164 is a blazar \citep[a subclass of radio-loud AGN with small viewing angles from the jet axis;][]{1980ARA&A..18..321A} at a redshift of $z = 0.94$ \citep{1987ApJ...318..577C}. Even among blazars, AO 0235+164 is found to be extremely compact in radio with no stable extended structure even at parsec-scale resolution \citep[e.g.,][]{2017ApJ...846...98J}. On kiloparsec scales, Very Large Array (VLA) images at 1.4~GHz \citep{2007ApJS..171..376C} and at 5~GHz \citep{2006PASJ...58..217F} reveal a weak, extended structure north-northwest of the compact core. This could be a consequence of the extremely small viewing angle found for this source \citep[e.g.,][Section~\ref{sec:VLBI-JetComp} of this paper]{2017ApJ...846...98J, 2018MNRAS.475.4994K}. A consequence of such a small viewing angle is the enhancement of the Doppler factor $\delta$ which governs relativistic effects such as Doppler boosting. A large $\delta$ could amplify even small variations in the source into detectable variability. AO 0235+164 is known to display variation in brightness at various timescales, from as short as a few hours in optical \citep{2004MNRAS.348..176S} to as long as years in radio \citep{2009AJ....137.5022N}. Quasi-periodicity was observed in X-rays with a short period of 17~days \citep{2009ApJ...696.2170R}. Longer periods of a few years were found in both long-term radio and optical data \citep[e.g.,][]{2001A&A...377..396R, 2021MNRAS.501.5997T}, hinting at the possible existence of jet precession. Space-Very Long Baseline Interferometry (Space-VLBI) observations reveal extreme brightness temperatures of up to $10^{14}$~K \citep{2000PASJ...52..975F, 2018MNRAS.475.4994K}, implying the presence of extreme physics within the jet.\\
\indent{}The previous flaring activity from 2006 to 2009 was extensively investigated during multiple campaigns \citep[e.g.,][]{2008A&A...480..339R, 2011ApJ...735L..10A, 2012ApJ...751..159A}. After a long period of quiescence, AO 0235+164 once again displayed multiple flares from radio to $\gamma$-rays from 2014 to 2018 (e.g., \citealt{2015ATel.7975....1C}; see also Figure~\ref{fig:gamma_radio_lc}). In this paper, we present our analysis of total flux measurements in the radio (15~GHz to 340~GHz) and $\gamma$-ray bands from 2013 to 2019 data collected from multiple monitoring programs. We investigate the jet parameters with high-resolution VLBI data at 43~GHz obtained with the Very Long Baseline Array (VLBA). We also estimate the magnetic field strengths within the jet from quasi-simultaneous radio spectra.\\
\indent{}We adopt a flat $\Lambda$CDM cosmology of $H_0 = 67.4\textrm{ km/s/Mpc}$, $\Omega_{\textrm{m}}=0.315$, and $\Omega_\Lambda=0.685$  \citep{2020A&A...641A...6P}. The luminosity distance at the redshift of AO 0235+164 is $D_\textrm{L} = 6303.5\textrm{ Mpc}$, and the linear scale is $8.12\textrm{ pc/mas}$, which was calculated with CosmoCalc \citep{2006PASP..118.1711W}.\footnote{\url{https://www.astro.ucla.edu/~wright/CosmoCalc.html}}

\section{Multiwavelength Data}\label{sec:mw-data}
\subsection{Radio and Submillimeter Data}\label{sec:mw-data-rd}
We collect single-dish and compact-array light-curve data for AO 0235+164 from a wide range of observing frequencies (15-340~GHz). We use data obtained from 2013 to 2019. We also collect complementary 43~GHz VLBA monitoring data of AO 0235+164. The details of the individual datasets are presented below.
\subsubsection{OVRO Monitoring of Fermi Blazars}\label{sec:mw-data-ovro}
AO 0235+164 is monitored by the 40 m telescope at the Owens Valley Radio Observatory (OVRO) as part of the 15~GHz monitoring program \citep{2011ApJS..194...29R}.\footnote{\url{https://sites.astro.caltech.edu/ovroblazars/}} Starting from late 2007, the OVRO 15~GHz monitoring program has regularly monitored the flux of $\sim$1800 AGNs with a typical flux density rms of 4 mJy and typical systematic errors of 3$\%$. In this paper, we use data obtained from 2013 January 4 (MJD 56296) to 2019 December 22 (MJD 58839). The average cadence for AO 0235+164 during this period is 7.5 days. The flux rms error during this period is typically two percent of the total flux density.

\subsubsection{Metsähovi 37~GHz Light Curve}\label{sec:mw-data-metsa}
The 14 m telescope at the Aalto University Metsähovi Radio Observatory (MRO) is used to monitor multiple AGNs and microquasars at a center frequency of 36.8~GHz. The detection limit of the telescope at this frequency is on the order of 0.2 Jy under optimal conditions. Data points with a signal-to-noise ratio (S/N) less than 4 are handled as non-detections. The flux density scale is set by observations of DR~21 with NGC~7027, 3C~274, and 3C~84 being used as secondary calibrators. Details of the data reduction and analysis may be found in \citet{1998A&AS..132..305T}. We use data obtained from 2013 January 11 (MJD 56303) to 2019 December 30 (MJD 58847). The average cadence for AO 0235+164 during this period is 8.5 days. The rms error during this period is typically nine percent of the total flux density.

\subsubsection{POLAMI Light Curve}\label{sec:mw-data-POLAMI}
The POLAMI (Polarimetric Monitoring of AGN at Millimetre Wavelengths) program \citep{2018MNRAS.473.1850A,2018MNRAS.474.1427A,2018MNRAS.473.2506T}\footnote{\url{http://polami.iaa.es}} is a long-term program to monitor the polarimetric properties (Stokes I, Q, U, and V) of a sample of around 40 bright AGN at 86 and 230\,GHz frequencies with the IRAM 30~m Telescope near Granada, Spain. The program has been running since October 2006, and it currently has a time sampling of $\sim$2 weeks. The XPOL polarimetric observing setup has been routinely used as described in \citet{2008PASP..120..777T} since the start of the program. The reduction and calibration of the POLAMI data presented here are described in detail in \citet{2010ApJS..189....1A, 2014A&A...566A..59A, 2018MNRAS.474.1427A}.

\subsubsection{ALMA Calibrator Database}\label{sec:mw-data-ALMA}
The Atacama Large Millimeter Array (ALMA) Calibrator Source Catalog\footnote{\url{https://almascience.eso.org/sc/}} is a database containing measurements of various compact radio sources for multiple frequency bands. The primary purpose of the database is to assist in selecting bandpass, phase, and flux calibrators for ALMA observations. AO 0235+164 has been observed since mid-2011 at frequencies of 91.5 and 103.5~GHz (both Band~3), 233~GHz (Band~6), and 343.5~GHz (Band~7). Observations at the two different frequencies in Band 3 are usually simultaneous. On days where observations were made only at a single frequency in Band 3, this was usually done at 91.5~GHz. Therefore, due to the close proximity of the two frequencies and the higher cadence of measurements at the lower frequency, we choose only the 91.5~GHz data for further evaluation. We use Band 3 data obtained from 2013 June 17 (MJD 56460) to 2019 December 31 (MJD 58848), Band 6 data obtained from 2013 December 21 (MJD 56647) to 2019 September 16 (MJD 58742), and Band 7 data obtained from 2013 July 1 (MJD 56474) to 2019 December 18 (MJD 58835). The average cadence at Band 3 is 11 days with a typical rms flux error of four percent of the total flux density. The average cadence at Band 6 is 75 days with a typical rms flux error of seven percent of the total flux density. The average cadence at Band 7 is 15 days with a typical rms flux error of nine percent of the total flux density.

\subsubsection{SMA Calibrator Database}\label{sec:mw-data-SMA}
The Submillimeter Array (SMA) is a connected array of eight 6 m dishes located on Maunakea, Hawaii. AO 0235+164 is included in an ongoing monitoring program at the SMA to determine the ﬂuxes of compact extragalactic radio sources that can be used as calibrators at mm wavelengths \citep{2007ASPC..375..234G}. Available potential calibrators are observed for 3 to 5 minutes, and the measured source signal strength is calibrated against known standards, typically solar system objects (Titan, Uranus, Neptune, or Callisto). Data from this program are updated regularly and are available at the SMA website.\footnote{\url{http://sma1.sma.hawaii.edu/callist/callist.html}} We use 226~GHz data obtained from 2013 January 8 (MJD 56300) to 2019 April 5 (MJD 58578) and 342~GHz data from 2013 February 13 (MJD 56336) to 2018 August 12 (MJD 58342). The average cadences of the 226~GHz and 342~GHz data are 24 days and 223 days, respectively. We note that half of the observations at 342~GHz are concentrated between 2016 January 2 (MJD 57389) and 2016 February 28 (MJD 57446). The rms flux error is typically seven percent of the total flux density at 226~GHz and ten percent of the total flux density at 342~GHz.

\subsubsection{VLBA 43~GHz Data}\label{sec:mw-VLBA-BU}
AO 0235+164 has been monitored by the VLBA at 43~GHz as part of the VLBA-BU-BLAZAR program \citep{2005AJ....130.1418J, 2016Galax...4...47J}\footnote{\url{https://www.bu.edu/blazars/VLBAproject.html}} since 2007 June 14 (MJD 54265). We use the calibrated VLBI data obtained between 2013 January 15 (MJD 56307) and 2019 October 19 (MJD 58775). We analyze the data by fitting elliptical and circular Gaussian model components to the calibrated visibility data using the MODELFIT procedure in Difmap \citep{1994BAAS...26..987S}.

\subsection{Gamma-ray Data}
\begin{figure*}
\includegraphics[width=0.9\linewidth]{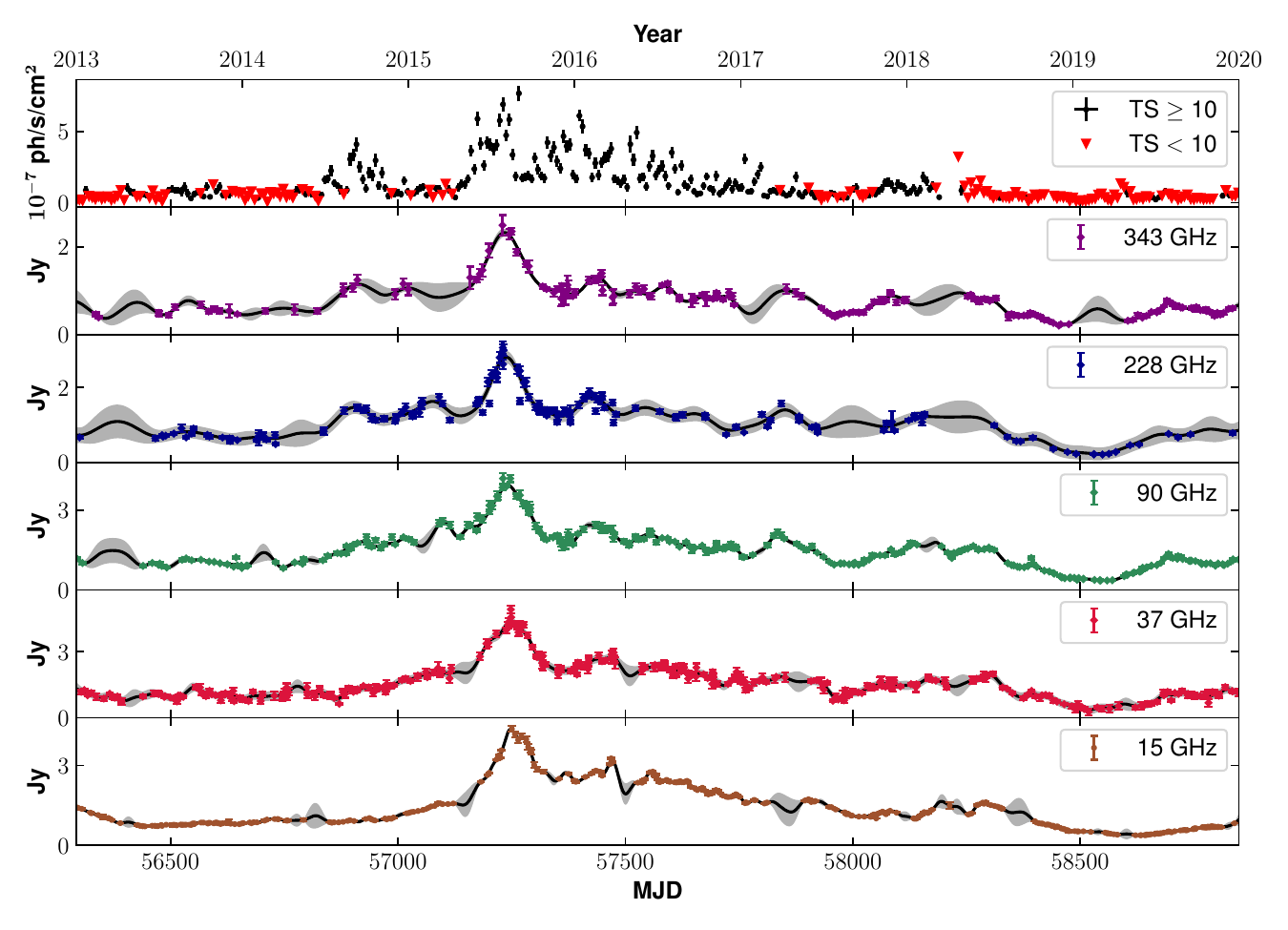}
\caption{Multiwavelength light curves of AO 0235+164 from 2013 to 2019. From top to bottom, the individual light curves correspond to $\gamma$-rays, 343~GHz, 228~GHz, 90~GHz, 37~GHz, and 15~GHz, respectively. Uncertainties correspond to 1$\sigma$ confidence regions. Additionally, the mean GPR light curves and corresponding 1$\sigma$ confidence regions for the radio data are plotted as black solid lines and grey shaded regions, respectively. In the $\gamma$-ray light curve, values with TS $\ge10$ are marked as black dots while upper-limits for time bins with TS $<10$ are plotted as red triangles.\label{fig:gamma_radio_lc}}
\end{figure*}

We use 100~MeV to 300~GeV $\gamma$-ray data obtained from the \textit{Fermi} Large Area Telescope \citep[LAT,][]{2009ApJ...697.1071A} between 2013 January 1 (MJD 56293) and 2019 December 31 (MJD 58848). We perform the data analysis using Fermitools (version 2.2.0) and the P8R3\_SOURCE\_V3 instrument response function. Photons in the event class 128 (`source' class) and within a region of interest (ROI) of 15 degrees centered at the position of AO 0235+164 were selected for the analysis. We carried out an unbinned likelihood analysis \citep{2009ApJS..183...46A} with a time bin of 7 days. All 4FGL \citep{2020ApJS..247...33A} sources within 25 degrees of AO 0235+164 were included in the initial xml model. We also include Galactic interstellar emission, using the \textit{gll\_iem\_v07.fits} model, and isotropic diffuse emission, using the \textit{iso\_P8R3\_SOURCE\_V3\_v1.txt} model\footnote{\url{https://fermi.gsfc.nasa.gov/ssc/data/access/lat/BackgroundModels.html}}. We also left free to vary the normalization of AO 0235+164, 16 variable sources (variability index $\ge$ 18.48) located within 10~degrees of AO 0235+164, and 4C +28.07, a $\gamma$-ray bright blazar located within 12~degrees of AO 0235+164. Model parameters of all other point sources were fixed. Spectral parameters for all sources were fixed to the 4FGL values. For AO 0235+164, we use a LogParabola function\footnote{\url{https://fermi.gsfc.nasa.gov/ssc/data/analysis/scitools/source_models.html\#LogParabola}} \citep{2006A&A...448..861M} to model the spectral shape of the $\gamma$-ray data.
For each 7-day time bin, we ran an initial likelihood analysis including all sources. After the initial run, we removed sources with a test statistic TS$<25$ from the xml model (with the exception of AO 0235+164). We then ran a second round of likelihood analysis with the filtered xml model. After the fit converged, we found the photon count rate (in units of ph~cm$^{-2}$~s$^{-1}$) along with the corresponding uncertainties using \textit{pyLikelihood}. The full $\gamma$-ray light curve from the start of 2013 to the end of 2019 is presented in Figure~\ref{fig:gamma_radio_lc}. In each time bin, we consider AO 0235+164 to be detected if the test statistic (TS) is $\ge10$ \citep[e.g.,][]{2012ApJ...751..159A}. 95$\%$ confidence upper limits are plotted for time bins with TS~$<10$.

\section{Data Analysis and Results}

\subsection{Multiwavelength Light Curve Correlations}\label{sec:MW_corr}
\begin{figure*}
\centering
\begin{subfigure}{0.49\textwidth}
    \centering
    \includegraphics[width=\textwidth]{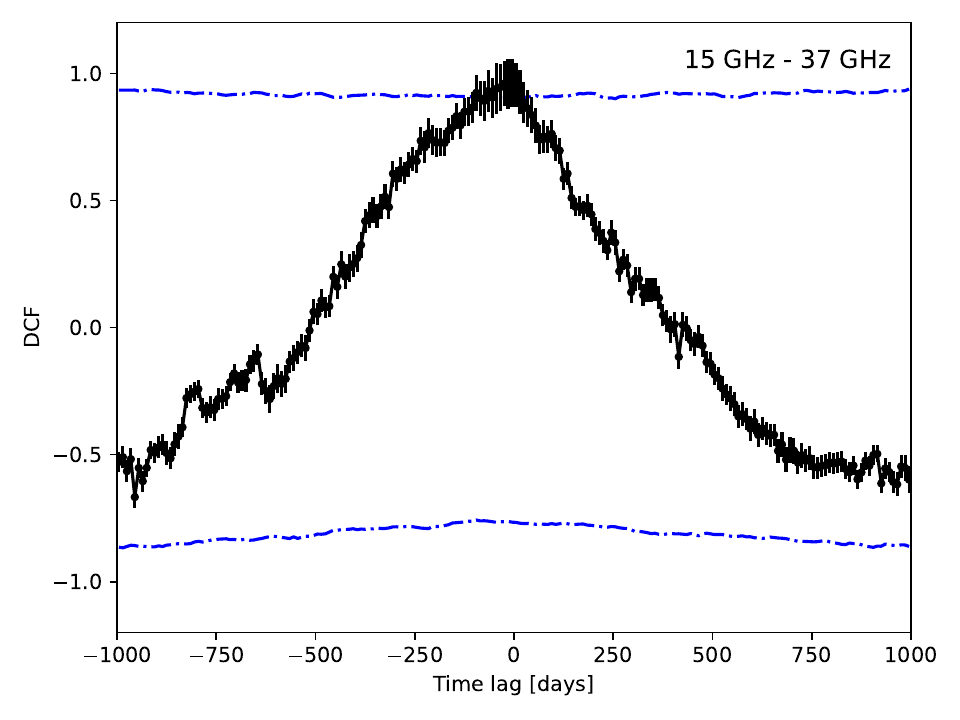}
\end{subfigure}
\begin{subfigure}{0.49\textwidth}
    \centering
    \includegraphics[width=\textwidth]{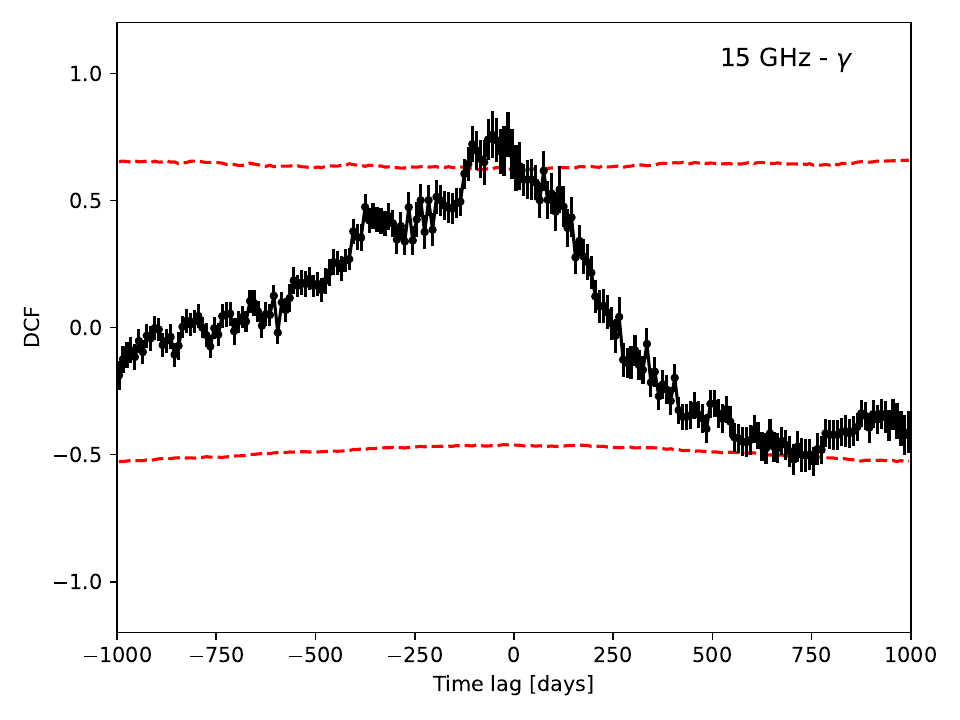}
\end{subfigure}
\\
\begin{subfigure}{0.33\textwidth}
    \centering
    \includegraphics[width=\textwidth]{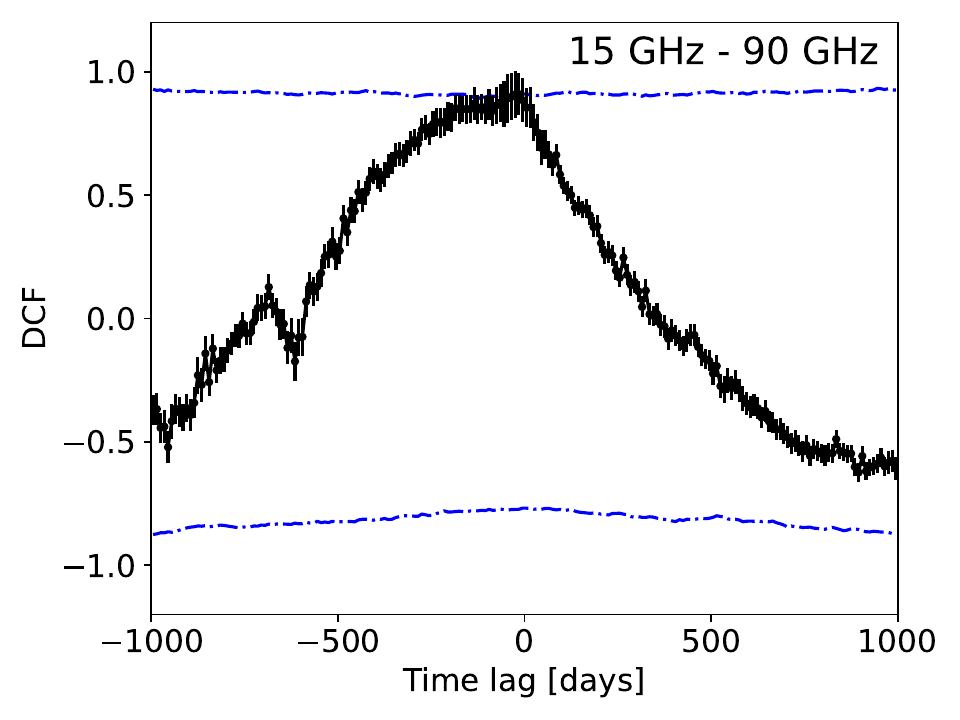}
\end{subfigure}
\begin{subfigure}{0.33\textwidth}
    \centering
    \includegraphics[width=\textwidth]{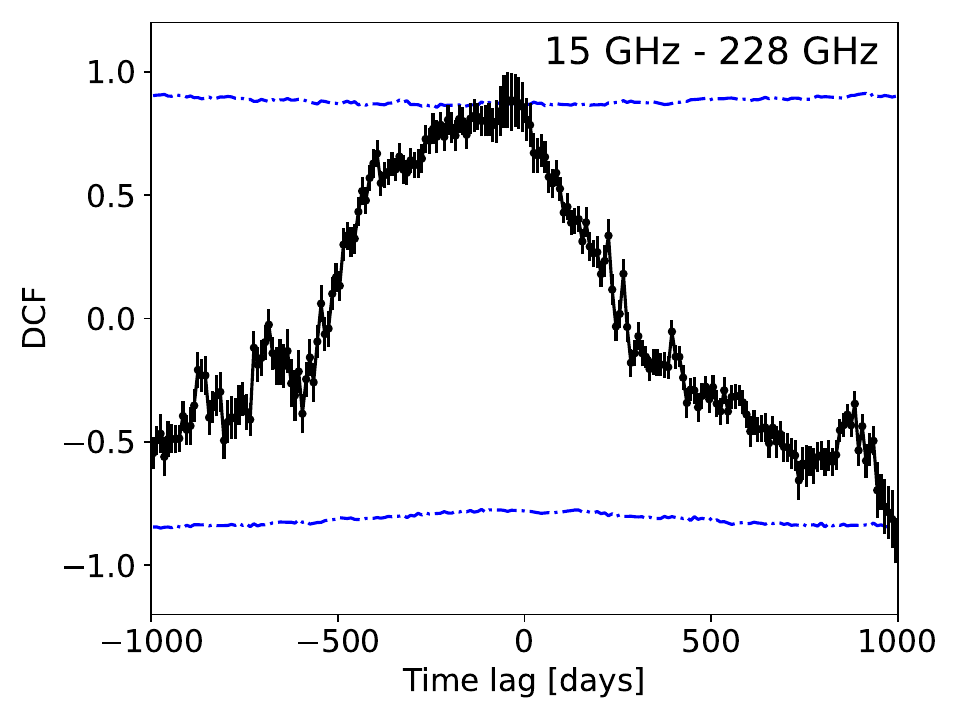}
\end{subfigure}
\begin{subfigure}{0.33\textwidth}
    \centering
    \includegraphics[width=\textwidth]{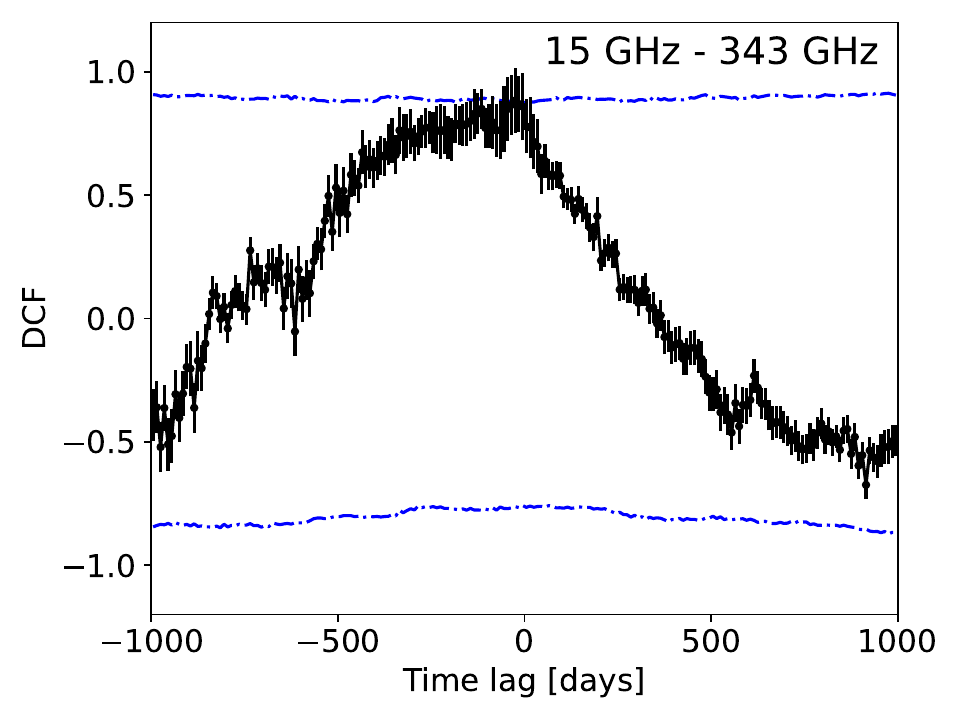}
\end{subfigure}
\caption{DCFs of the 37~GHz, 90~GHz, 228~GHz, 343~GHz, and $\gamma$-ray light curves with respect to the 15~GHz light curve. The specific light curve pair is labeled in the upper-right corner of each plot. Negative time lags indicate that the paired light curve leads the 15~GHz light curve. The red dashed horizontal lines represent the boundaries of the $95.45\%$ confidence region. The blue dot-dashed horizontal lines represent the boundaries of the $99.73\%$ confidence region.}
\label{fig:dcf_full}
\end{figure*}

We search for correlations among the various light curves using the discrete correlation function (hereafter DCF), which is a method for analyzing correlations in unevenly sampled data sets. Details on the DCF may be found in \citet{1988ApJ...333..646E}. To improve the cadences of the light curves, we integrate data sets taken at similar frequencies into combined light curves. We do this for the following data sets: (1) we combine the POLAMI 86~GHz data and ALMA~91.5 GHz data into the 90~GHz light curve; (2) we combine the POLAMI~229 GHz data, the ALMA 233~GHz data, and the SMA~226 GHz data into the 228~GHz light curve; and (3) we combine the ALMA 343.5~GHz data and the SMA 342~GHz into the 343~GHz light curve. The mean frequencies of the combined light curves are 90.4~GHz, 228~GHz, and 343~GHz respectively. 
We evaluate DCFs for all of the light curves with respect to the 15~GHz light curve, which has both a high mean cadence and low measurement errors. The 37~GHz light curve has a comparable mean cadence, but the measurement errors are significantly larger.\\
\indent{}To evaluate the uncertainties of the measured time lags $\tau_{\rm DCF}$, we simulate $10^4$ artificial light curves using the Flux Randomization (FR) and Random Subset Selection (RSS) methods described in \citet{1998PASP..110..660P}. For each frequency pair, we fit a normalized distribution to the array of $10^4$ simulated DCF peaks ($A_{\textrm{DCF}}$) and corresponding time lags. The fit parameters are presented in Table~\ref{tab:Multiwave_DCF_TB}. We note that $\tau_{\rm DCF}<0$ indicates that the corresponding light curve is leading the 15~GHz light curve. We find that all radio light curves from 37~GHz to 343~GHz have $\tau_{\rm DCF}<0$, indicating that flares at higher frequencies occur prior to those at lower frequencies. There is a trend of larger absolute values of $\tau_{\rm DCF}$ with increasing frequency of the reference light-curve, although this is difficult to track concretely due to the relatively large uncertainties of $\tau_{\rm DCF}$. We find that the $\gamma$-rays also lead the radio frequencies, indicating that the $\gamma$-ray region is located upstream of the radio emission region (see Section~\ref{sec:equpartTest} for additional analysis).\\
\indent{}We evaluate the significance of the correlation by simulating $10^4$ light curves using the method of \citet{2013MNRAS.433..907E} as implemented in DELCgen \citep{2016ascl.soft02012C}. The probability density function (PDF) and the power spectral density (PSD) encapsulate the variability and statistical properties of a given light curve, and it has been demonstrated that both properties used for the creation of simulated light curves may significantly affect the estimation of the confidence intervals of a DCF \citep[e.g.,][]{2013MNRAS.433..907E,2014MNRAS.445..437M}. Therefore, it is important that the simulated light curves accurately represent both the PDF and PSD of the original light curve. We model the PDFs of the radio light curves using a log-normal distribution \citep[e.g.,][]{2020MNRAS.494.3432G,2021MNRAS.504.1427A}, whereas we use a gamma distribution for the $\gamma$-ray light curve \citep[e.g.,][]{2018ApJ...852...30A,2022ApJ...925...64K}. We model the PSDs for all frequencies using a power-law function with an added constant \citep[e.g.,][]{2014ApJ...785...76P, 2020ApJS..250....1T}.\\
\indent{}Examples of the calculated DCFs with the corresponding confidence intervals are presented in Figure~\ref{fig:dcf_full}. All five light curves, from 37~GHz to $\gamma$-rays, are correlated with the 15~GHz light curve with the significance of the correlation with the $\gamma$-rays reaching a level of greater than 2$\sigma$. AO 0235+164 is known for significant correlations between radio and $\gamma$-ray flares \citep[e.g.,][]{2014MNRAS.445..428M}. We find that this is the case for the period of our study, suggesting that radio and $\gamma$-ray emission are tightly related in this source. The 3$\sigma$ correlations between the radio data sets suggest that the radio fluxes of this source arise from a common origin and mechanism.

\begin{table}
\caption{Fit parameters of the DCFs computed with respect to the 15~GHz light curve. Column (1) is the frequency of the companion light curve with the exception of the $\gamma$-ray data*. Column (2) is the peak of the DCF. Column (3) is the time lag. Quoted uncertainties represent 1$\sigma$ confidence regions on the parameters}
\label{tab:Multiwave_DCF_TB}
\begin{tabular}{ccc}
\hline
Frequency* & $A_{\textrm{DCF}}$ & $\tau_{\textrm{DCF}}$ \\
$\textrm{[GHz]}$ &  & [days]\\
\hline
$\gamma$-ray& $0.74 \pm 0.04$ & $-48 \pm 29$ \\
343 & $0.87 \pm 0.02$ & $-24 \pm 10$\\
228 & $0.89 \pm 0.02$ & $-37 \pm 15$\\
90 & $0.90 \pm 0.01$ & $-24 \pm 12$\\
37 & $0.95 \pm 0.01$ & $-17 \pm  9\ \ $\\
\hline
\end{tabular}
\end{table}

\subsection{Radio Light Curve Gaussian Process Regression}
Given the variety of multiwavelength observations used in this analysis, our measurements at different wavelengths are not simultaneous. To construct quasi-simultaneous radio spectra, we interpolate the data to a common date with Gaussian Process Regression \citep[hereafter GPR,][]{2006gpml.book.....R}, which has been used in the past to study the variability characteristics of blazars \citep[e.g.,][]{2016A&A...590A..48K}. We use \textit{Scikit-learn} \citep{scikit-learn} for the GPR, using a noisy radial-basis function (RBF) kernel of the form
\begin{equation}
k_\textrm{n}\left(x_\textrm{i},x_\textrm{j}\right)=\sigma^2_\textrm{f}\exp\left[-\frac{1}{2}\left(\frac{x_\textrm{i}-x_\textrm{j}}{l}\right)^2\right]+\sigma^2_{\textrm{rms}}\delta_{\textrm{ij}}
\end{equation}
to represent the covariance between two data points at $x_\textrm{i}$ and $x_\textrm{j}$. There are a total of three hyperparameters. Parameters $\textit{l}$ and $\sigma_\textrm{f}$ represent the timescale and magnitude of the covariation, respectively. The parameter $\sigma_{\textrm{rms}}$ corresponds to a global uncertainty due to the measurement errors. Together with the Kronecker delta function $\delta_{\textrm{ij}}$, $\sigma_{\textrm{rms}}$ is part of the white kernel, $\sigma^2_{\textrm{rms}}\delta_{\textrm{ij}}$. The hyperparameters are optimized separately for the 15, 37, 90, 228, and 343~GHz radio light curves using a maximum likelihood analysis. After optimization, the mean GPR light curve is obtained for 1-day intervals starting from 2013 January 1 (MJD 56293). We plot the results with the observations in Figure~\ref{fig:gamma_radio_lc} using black solid lines and grey shaded regions to indicate GPR light curves and their corresponding 1$\sigma$ confidence regions.

\subsection{Radio Spectral Energy Distribution}\label{sec:radio_curve_spectrum}
\begin{figure}
\includegraphics[width=\columnwidth]{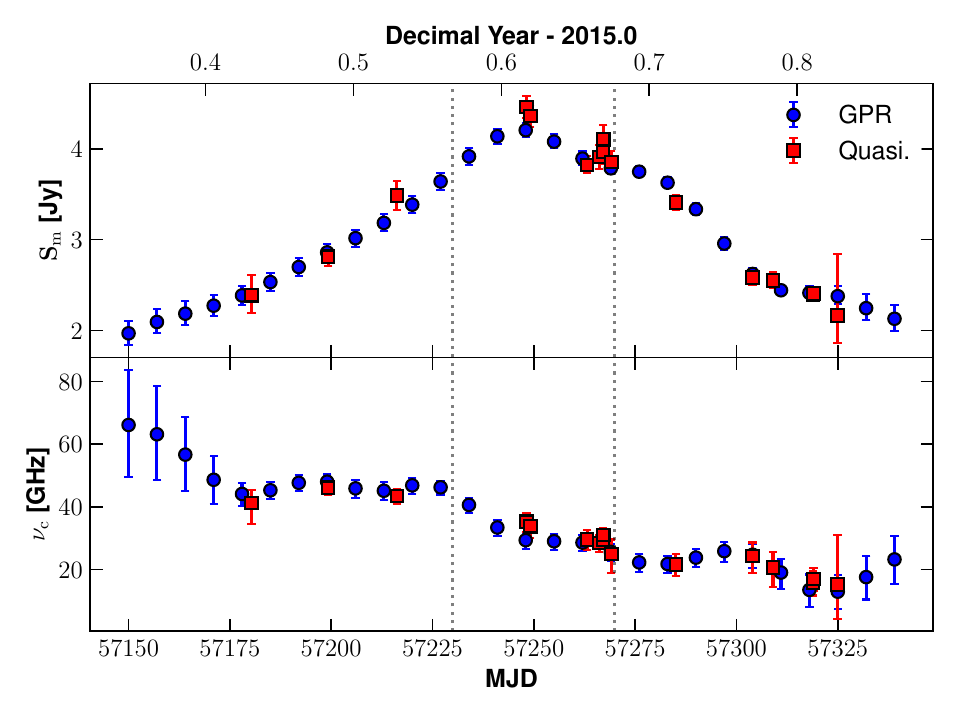}
\caption{Evolution of $S_{\rm m}$ and $\nu_{\rm c}$ during the 2015 flare. Quasi-simultaneous spectrum results are in red squares while GPR spectrum results are in blue circles. The vertical dotted lines indicate the boundaries of the different phases described in Sec.~\ref{sec:radio_curve_spectrum}. The time gap between consecutive GPR data points is 7 days (in order to prevent overlap). \label{fig:turnover_timecurve}}
\end{figure}

\begin{figure*}
\centering
\begin{subfigure}{0.49\textwidth}
    \centering
    \includegraphics[width=\textwidth]{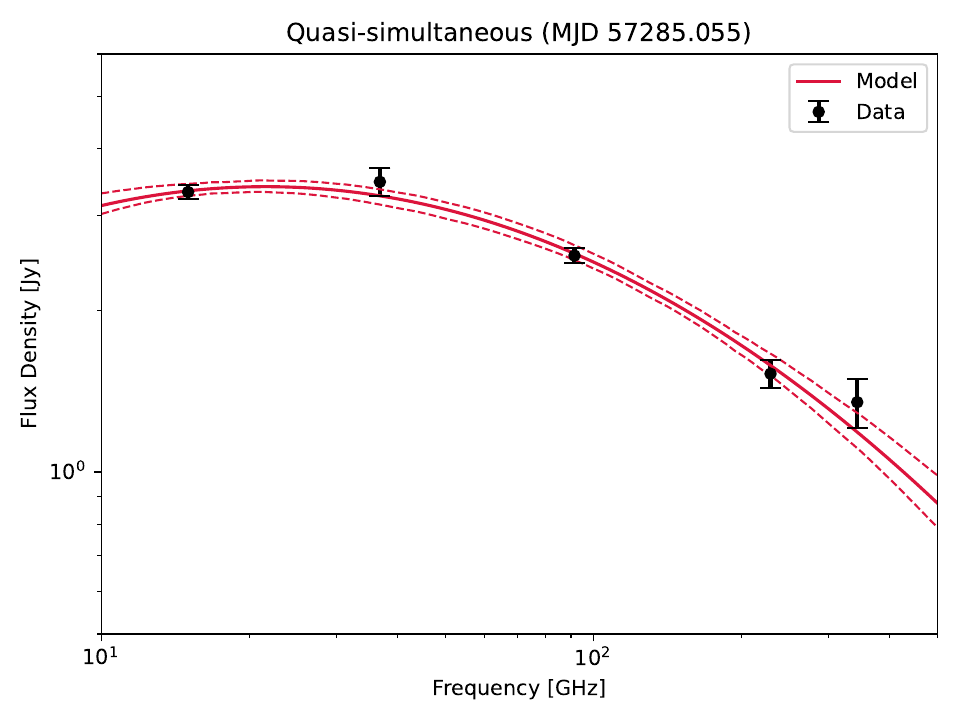}
\end{subfigure}
\begin{subfigure}{0.49\textwidth}
    \centering
    \includegraphics[width=\textwidth]{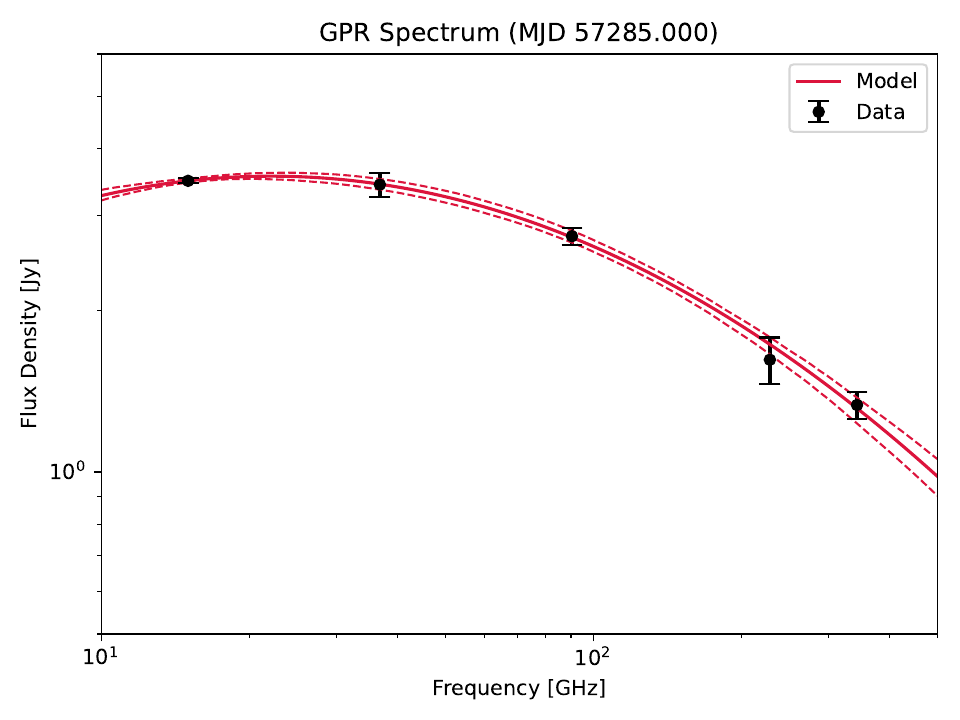}
\end{subfigure}
\\
\begin{subfigure}{0.49\textwidth}
    \centering
    \includegraphics[width=\textwidth]{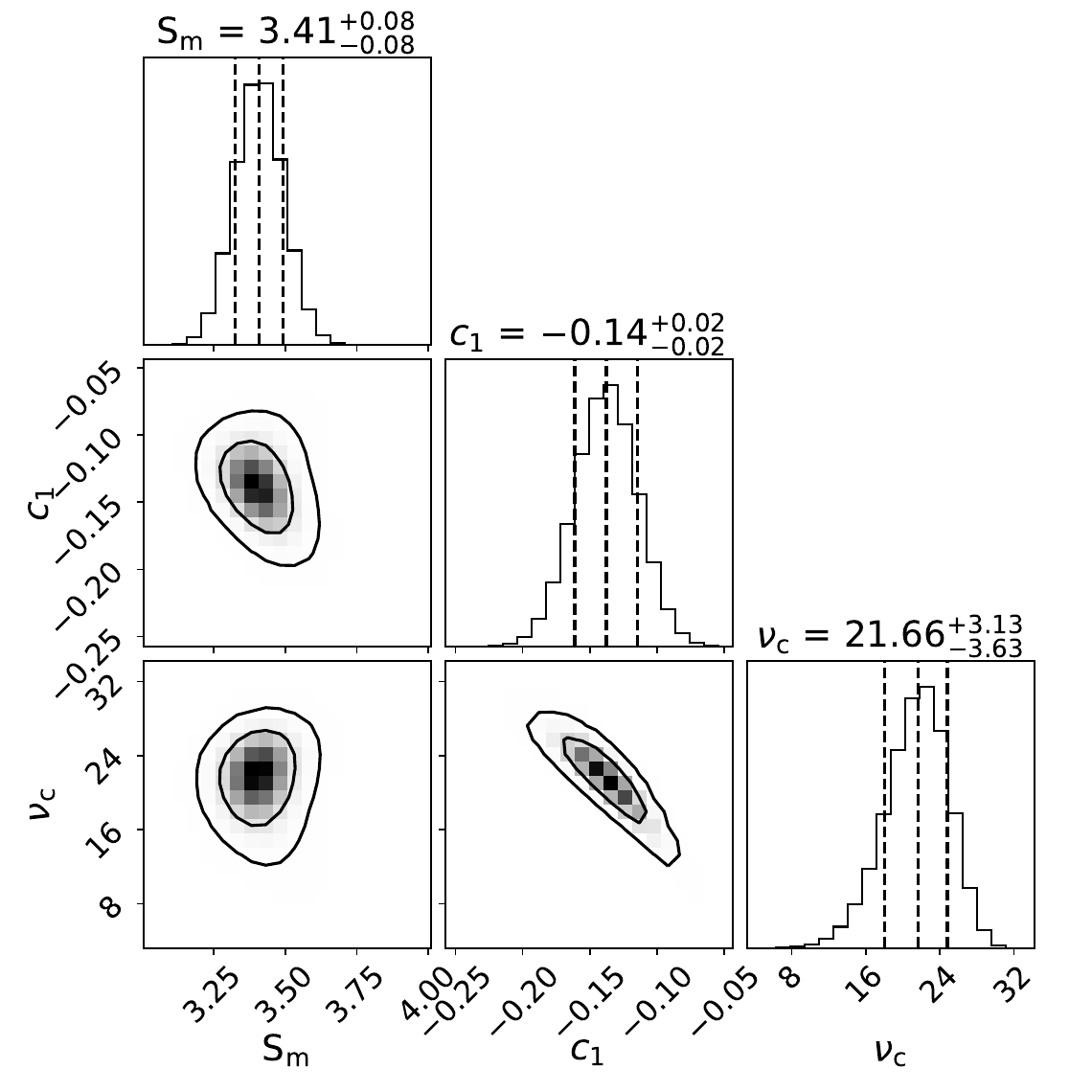}
\end{subfigure}
\begin{subfigure}{0.49\textwidth}
    \centering
    \includegraphics[width=\textwidth]{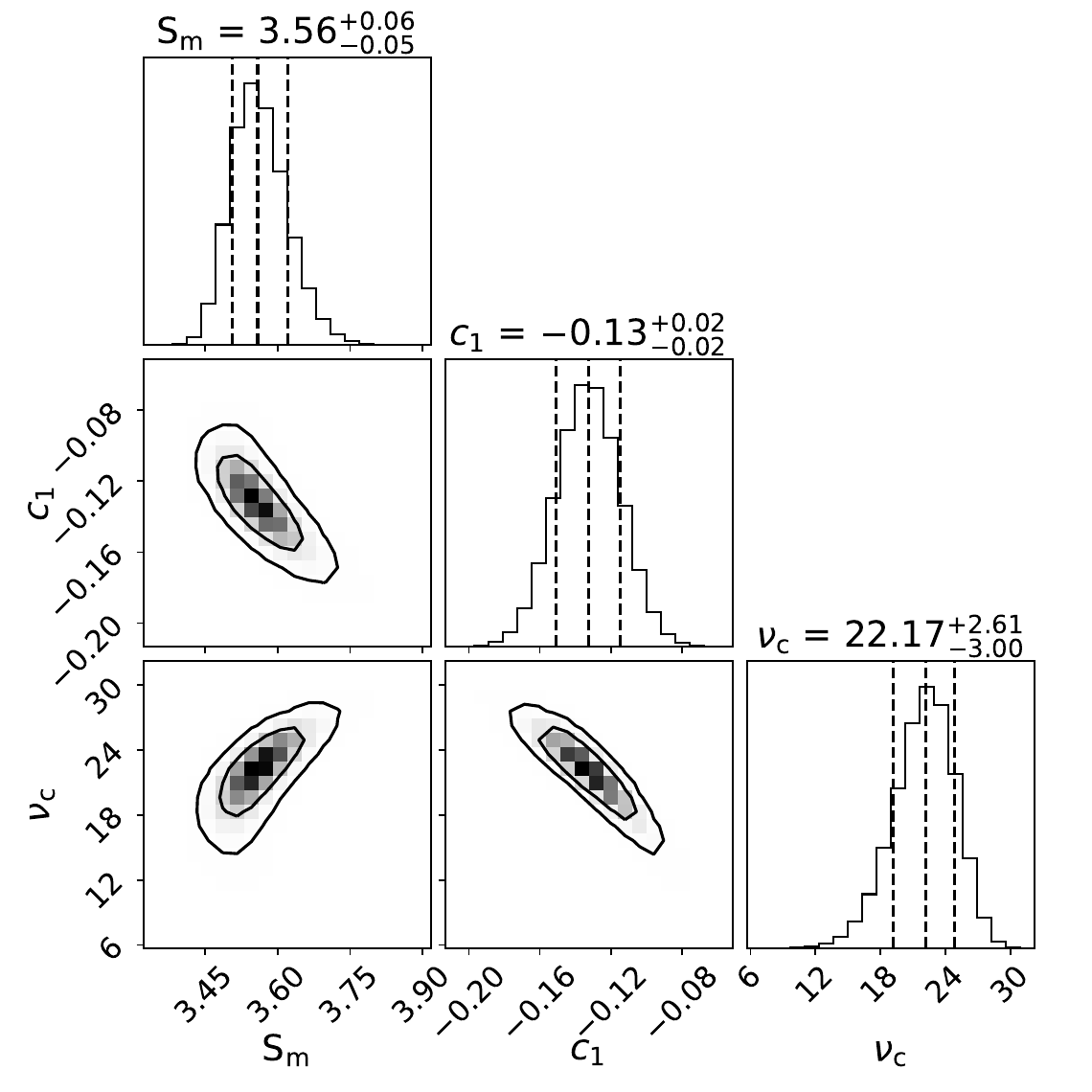}
\end{subfigure}
\caption{Example of a spectral fit for the spectrum during the flare occurring on 2015 September 20 (MJD 57285). The upper row plots the radio spectrum of the source in black circles with the derived spectral model in red. The lower row shows the joint posterior distribution of the fit parameters. Results for the quasi-simultaneous spectrum are plotted in the left column while those corresponding to the GPR spectrum are plotted in the right.}
\label{fig:ssa_spec_example}
\end{figure*}

\begin{figure}
\includegraphics[width=\columnwidth]{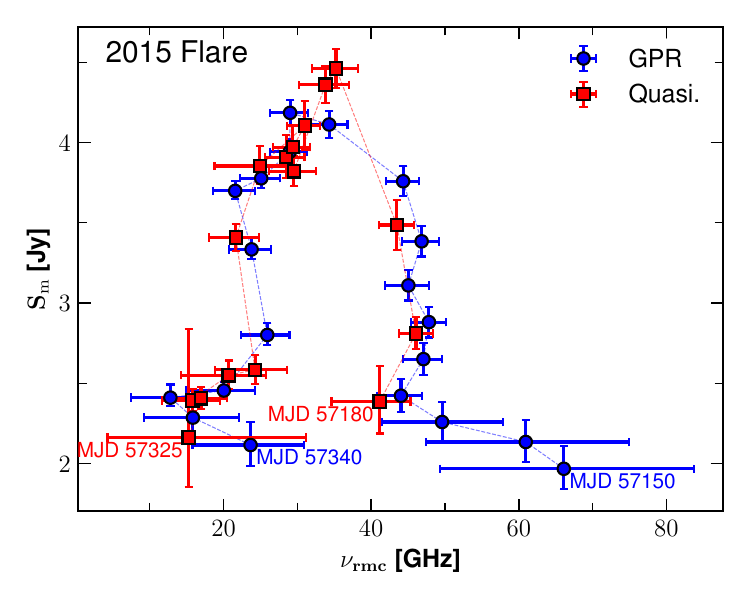}
\caption{Plot of $S_{\rm m}$ vs $\nu_{\rm c}$ derived from the GPR spectrum (blue circles) and the quasi-simultaneous spectrum (red squares) during the 2015 flare. The GPR data start at MJD 57150 and end at MJD 57340. The time gap between consecutive GPR data points is 10 days (in order to prevent overlap). The quasi-simultaneous data start at MJD 57180 and end at MJD 57325.\label{fig:turnover_param_2015}}
\end{figure}

We investigate spectral evolution in the radio for AO 0235+164 during its 2015 to 2019 flaring period. We limit our evaluation to periods when the total radio flux is dominated ($>90$\%) by the radio core as inferred from the 43~GHz VLBA data (see discussion in Section~\ref{sec:SD-Core_Dominance}). There are two periods that satisfy this criterion---the flaring activity in 2015 March-November (MJD~57100-57350) and the flaring activity in 2018 March-August (MJD~58180-58360). We fit a curved power law of the form \citep[e.g.,][]{2006A&A...448..861M, 2016ApJS..227....8L, 2018ApJ...859..128A}
\begin{equation}
S\left(\nu\right)=S_{\textrm{m}}\left(\frac{\nu}{\nu_\textrm{c}}\right)^{c_1\ln\left(\frac{\nu}{\nu_\textrm{c}}\right)}
\end{equation}
to the single dish and compact array radio light curves for these two periods. $S_{\rm m}$ is the flux at the turnover frequency $\nu_{\rm c}$. $c_1$ is a constant that fits the overall spectral shape. We fit to both the quasi-simultaneous spectrum, which was constructed with observations within 3 days of Metsähovi 37~GHz observations, and the GPR spectrum constructed from the GPR light curves. In both cases, we remove the minimum detected flux from each frequency as a method of exploring the variable radio flux. The evolution of the resultant parameters of the spectral analysis for the 2015 flaring activity is plotted in Figure~\ref{fig:turnover_timecurve}. $S_{\rm m}$ and $\nu_{\rm c}$ values from the quasi-simultaneous spectrum are plotted in red while those from the GPR spectrum are plotted in blue. The uncertainties correspond to the 1$\sigma$ confidence regions for the parameters which are estimated with a Markov Chain Monte Carlo (MCMC) analysis. We present an example of a fit in Figure~\ref{fig:ssa_spec_example}. For this example, we see that spectral shapes (characterized by $c_1$ and $\nu_{\rm c}$) determined with the two different methods are well compatible. The extimated values of $S_{\rm m}$ differ by $\sim5\%$.
Overall, we find that the evolution of the spectral parameters ($S_{\rm m}$ and $\nu_{\rm c}$) is well interpolated by using the GPR light curves. We present $S_{\rm m}$ with respect to $\nu_{\rm c}$ as derived from the GPR spectrum for the 2015 flare in Figure~\ref{fig:turnover_param_2015}. The evolution of the flare may be decomposed into three stages. During the onset of the flare, $S_{\rm m}$ rises from $\sim$2~Jy to $\sim$3.6~Jy while $\nu_{\rm c}$ is approximately constant at 45~GHz. Then, $\nu_{\rm c}$ decreases from $\sim$44~GHz at MJD~57230 to $\sim$25~GHz at MJD~57270. During this period, $S_{\rm m}$ increases from $\sim$3.6~Jy to $\sim$4.4~Jy, followed by a decrease to $\sim$3.7~Jy. Finally, $S_{\rm m}$ decreases from $\sim$3.7~Jy to $\sim$2~Jy as $\nu_{\rm c}$ is $\sim20$~GHz.

\subsection{Jet Parameters from New Superluminal Jet Components}\label{sec:VLBI-JetComp}
\begin{figure*}
\includegraphics[width=0.7\textwidth]{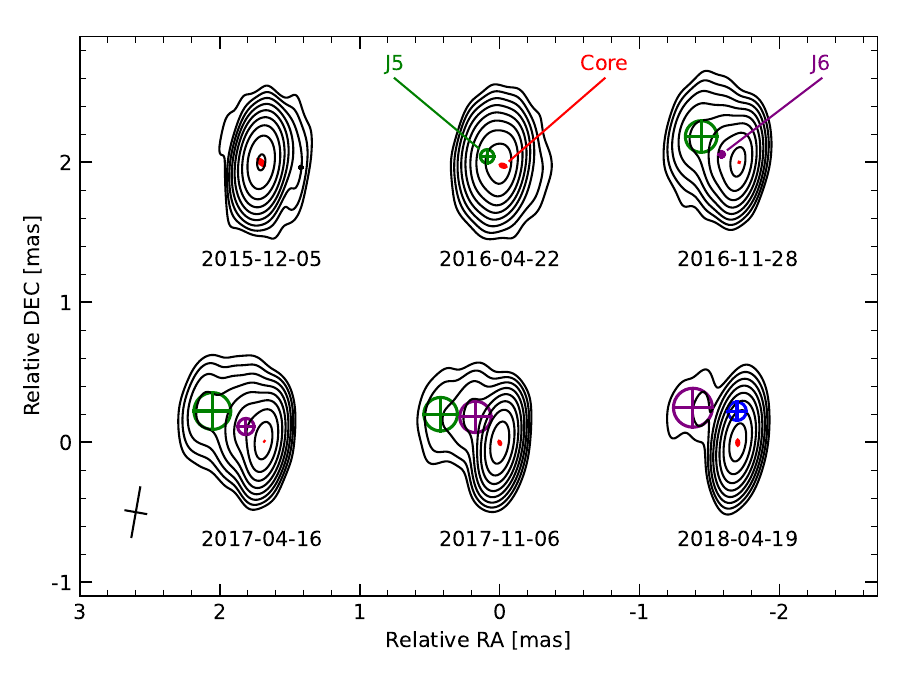}
\caption{A subset of the 43~GHz VLBA total intensity images of AO 0235+164 following the peak of the 2015 flare. All CLEAN maps were restored with a common beam of $0.36\times0.15\textrm{ mas}$ and the beam major axis at a position angle of $-10$ degrees (plotted in the lower-left corner). The contours start at $5\textrm{ mJy/beam}$ and incrementally increase by factors of 2. The images from each epoch are shifted in R.A. and Dec. to avoid overlap. The model-fitted components are overlaid on top of the CLEAN maps. The radio core (red) is the compact feature located near the intensity peaks. Jet components are plotted in $\bigoplus$ symbols with different colors used to identify different components. The J5 and J6 components studied in Section~\ref{sec:VLBI-JetComp} are plotted in green and purple, respectively.}
\label{fig:7mmVLBI_Maps}
\end{figure*}

\begin{figure}
\includegraphics[width=\columnwidth]{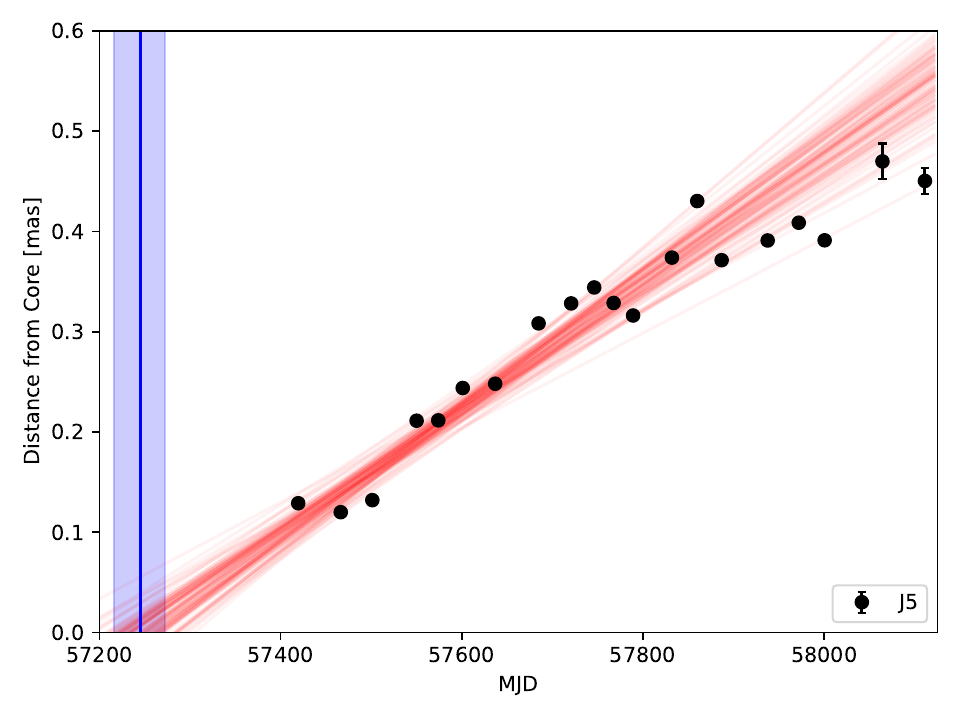}
\caption{The separation between the newly ejected jet component J5 (associated with the 2015 multiwavelength flare) and the radio core. This component was ejected from the 43~GHz core on MJD $57246^{+26}_{-30}$ (plotted in blue) with a proper motion of $\mu=0.230\pm0.019\textrm{ mas/yr}$ (plotted in red).
	\label{fig:VLBICompSep}}
\end{figure}

\begin{figure}
\includegraphics[width=\columnwidth]{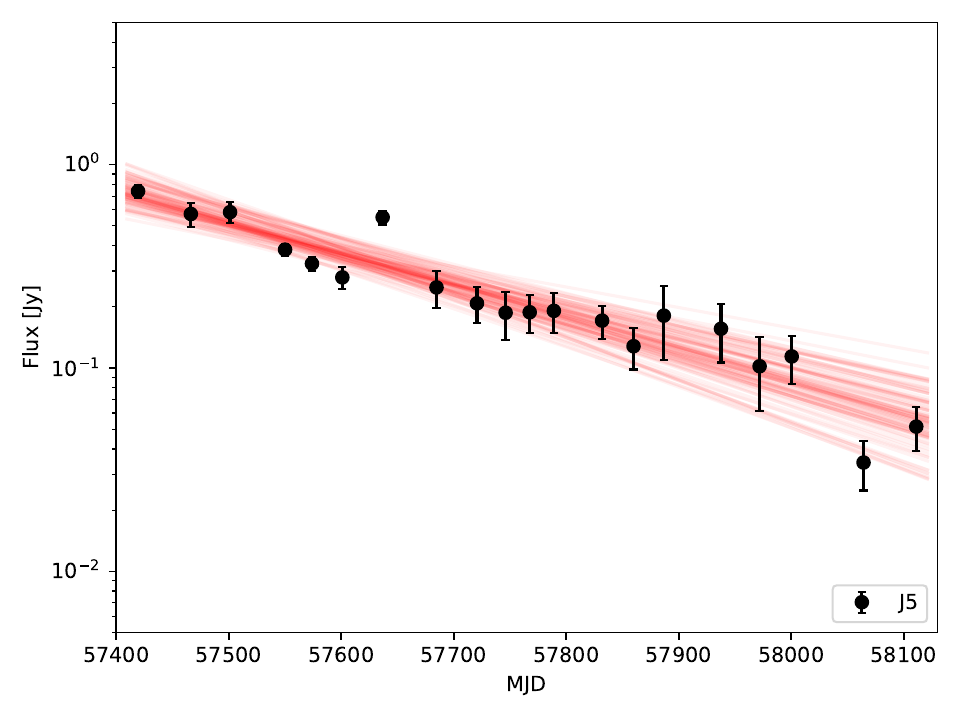}
\caption{The decay of the total flux of J5. Fitting an exponential decay resulted in a variability timescale of $\tau_{\rm var}=0.76^{+0.14}_{-0.10}\textrm{ yr}$.
	\label{fig:VLBICompFlux}}
\end{figure}

While a detailed analysis of the VLBI data is beyond the scope of this paper, we present a subset of our study on determining the jet parameters of AO 0235+164 during the 2015 flare (which is required for the following analysis presented in this paper). We conducted model fitting of the calibrated 43~GHz VLBI data from the VLBA-BU-BLAZAR program with Difmap. We model the radio core with an elliptical Gaussian (i.e., a 2D elliptical Gaussian brightness distribution) and any other components with circular Gaussians. The positions of all components, including the core, were set as free parameters. We evaluate the errors in the fit parameters following the procedures in \citet{1999ASPC..180..301F} and \citet{2008AJ....136..159L}. From the 43~GHz VLBA data, we find the ejection of a new jet component (hereafter J5) at a position angle of approximately $57.8\pm1.7\textrm{ degrees}$ (see Figure~\ref{fig:7mmVLBI_Maps}). We were able to track this component up to approximately $0.4\textrm{ mas}$ from the core before it became too faint to be reliably detected. The time evolution of the distance between J5 and the core is shown in Figure~\ref{fig:VLBICompSep}. Fitting a linear function to the separation (plotted in red), we find a proper motion of $\mu=0.230\pm0.019\textrm{ mas/yr}$ and a corresponding apparent speed of $\beta_{\rm app}=11.84^{+0.96}_{-0.97} c$. The epoch of ejection from the core is estimated to be  $\textrm{MJD }57246^{+26}_{-30}$, which corresponds to the peak of the 2015 flare. \\
\indent{}The light curve of J5 is presented in Figure~\ref{fig:VLBICompFlux}. Following \citet{2017ApJ...846...98J}, we estimate the variability Doppler factor from the decaying light curve of this jet component. We model the light curve as an exponential decay with the function $\ln{\left(S(t)/S_0\right)}=-k_{\rm var}\left(t-t_{0}\right)$, where $S_0$ is the flux of the component at a certain time $t_{0}$ and $k_{\rm var}$ is a decay constant. The variability timescale is then found as $\tau_{\rm var}=\left|1/k_{\rm var}\right|$. Fitting to the data of J5, we find $k_{\rm var}=1.32\pm0.20\textrm{ yr}^{-1}$ and $\tau_{\rm var}=0.76^{+0.14}_{-0.10}\textrm{ yr}$. Following \citet{2005astro.ph..3225L}, we calculate the minimum resolvable size as $\theta_{\rm min}=2^{1-\beta/2}b_{\psi}\sqrt{\frac{\ln{2}}{\pi}\ln{\left(\frac{\textrm{S/N}}{\textrm{S/N}-1}\right)}}$, where $b_\psi$ is the full width at half maximum (FWHM) at an arbitrary position angle $\psi$, and $\beta$ is a weighting parameter with $\beta=0$ for uniform weighting and $\beta=2$ for natural weighting. If the fit FWHM was found to be smaller than $\theta_{\rm min}$, we consider the component to be unresolved. The mean FWHM of J5 was found to be $\left<\theta_{\rm vlbi}\right>_{\rm J5}=0.234\pm0.059\textrm{ mas}$, which was calculated over a total of 9 epochs where the component was found to be resolved. The variability Doppler factor is found as \citep{2017ApJ...846...98J}
\begin{equation}
\delta_{\rm var}=\frac{15.8}{1+z}\left[\frac{d_{\rm m}}{1\textrm{ mas}}\right]\left[\frac{D_{\rm L}}{1\textrm{ Gpc}}\right]\left[\frac{\left|k_{\rm var}\right|}{1\textrm{ yr}^{-1}}\right] = 28.5 \pm 8.4,
\end{equation}
where we have substituted $\tau_{\rm var}^{-1}$ with $\left|k_{\rm var}\right|$ and estimated the angular size assuming a uniform sphere (i.e., 1.8 times $\theta_{\rm VLBI}$ for consistency with Section~\ref{sec:bssabeq}) rather than the uniform disk assumed in \citet{2017ApJ...846...98J}. This difference does not significantly affect the derived jet parameters.\\
\indent{}With an independent estimate of $\delta_{\rm var}$ and $\beta_{\rm app}$, we are now able to determine the bulk Lorentz factor $\Gamma$ and the intrinsic viewing angle $\theta_{\rm v}$ as \citep[e.g.,][]{2012rjag.book.....B}
\begin{equation}
\Gamma = \frac{\beta_{\rm app}^{2}+\delta_{\rm var}^2+1}{2\delta_{\rm var}}=16.8^{+3.6}_{-3.1}
\end{equation}
and
\begin{equation}
\tan{\theta_{\rm v}}=\frac{2\beta_{\rm app}}{\beta_{\rm app}^{2}+\delta_{\rm var}^2-1}=0.025^{+0.019}_{-0.009}.
\end{equation}
This corresponds to an intrinsic viewing angle of $\theta_{\rm v}=1.42^{+1.07}_{-0.52}\textrm{ degrees}$. We note that the values of $\delta_{\rm var}$ and $\Gamma$ for J5 are significantly smaller than those presented in \citet{2017ApJ...846...98J} for the 2008 flare ($\delta\approx60$, $\Gamma\approx31$). However, \citet{2018MNRAS.475.4994K} report $\delta\approx24$ and $\Gamma\approx14$ for the same 2008 flare. \citet{2009A&A...494..527H} also find $\delta=24$ and $\Gamma=12.1$ for this source, which were derived from the variability timescales of long-term radio monitoring data. Both \citet{2009A&A...494..527H} and \citet{2018MNRAS.475.4994K} are consistent with the values found in this paper for the 2015 flare. \\
\indent{}We find the presence of a second component (hereafter J6) ejected close in time to J5. Following the same analysis, we find that J6 has a slower proper motion, likely due to a smaller viewing angle as compared to J5. The value of $\delta_{\rm var}$ is consistent with that of J5 within uncertainties while the estimated values of $\Gamma$ agree well between the two jet components. The results are summarized in Table~\ref{tab:vlbijetparam}. \citet{2022ApJS..260...12W} analyze the same 43~GHz VLBI data during this period and also find the two jet components J5 and J6 in Table~\ref{tab:vlbijetparam} (labeled respectively as B5 and B6 in their paper). The position angles of the jet components found in \citet{2022ApJS..260...12W} agree well with those found in our analysis; on the other hand, the proper motion of J5 is $\sim20\%$ slower and that of J6 is $\sim60\%$ faster than in Table~\ref{tab:vlbijetparam}. The variability timescales in \citet{2022ApJS..260...12W} are $\sim50\%$ lower than found in this paper. Their $\delta_{\rm var}$ and $\Gamma$ for J5 are consistent with our results while those for J6 differ by approximately a factor of two.\\
\indent{}Due to J5 having a faster, well-constrained proper motion compared to J6, we consider the jet parameters derived from J5 to be representative of AO 0235+164 at the time of the 2015 flare. We refer to these values in the following analysis.
\begin{table*}
\caption{Jet parameters of AO 0235+164 derived from the jet components J5 and J6. The columns correspond to (1) the ID of the jet component, (2) the epoch of ejection from the core, (3) the proper motion, (4) the direction of proper motion, (5) the apparent speed, (6) the variability timescale, (7) the mean size, (8) the variability Doppler factor, (9) the bulk Lorentz factor, and (10) the intrinsic viewing angle. All uncertainties correspond to 1$\sigma$ confidence regions for the parameters.}
\label{tab:vlbijetparam}
\renewcommand{\arraystretch}{1.5}
\begin{tabular}{cccccccccc}
\hline
ID& T$_{\rm o}$ & $\mu$ & PA & $\beta_{\rm app}$ & $\tau_{\rm var}$ & $\left<\theta_{\rm vlbi}\right>$ & $\delta_{\rm var}$ & $\Gamma$ & $\theta_{\rm v}$ \\
& [MJD] & [mas/yr] & [deg] & [c] & [yr] & [mas] &&&[deg]\\
\hline
J5 & $57246^{+26}_{-30}$ & $0.230\pm0.019$ & $57.8\pm1.7$ & $11.84^{+0.96}_{-0.97}$ & $0.76^{+0.14}_{-0.10}$ & $0.234\pm0.059$ & $28.5\pm8.4$ & $16.8^{+3.6}_{-3.1}$ & $1.42^{+1.07}_{-0.52}$\\
J6 & $56985^{+203}_{-388}$ & $0.073\pm0.023$ & $50.7\pm1.4$ & $3.74^{+1.20}_{-1.18}$ & $0.87^{+0.22}_{-0.22}$ & $0.330\pm0.097$ & $35.0\pm13.6$ & $17.8^{+6.6}_{-6.8}$ & $0.36^{+0.64}_{-0.20}$\\
\hline
\end{tabular}
\end{table*}

\subsection{Radio Core Magnetic Field Strength}\label{sec:bssabeq}
\begin{figure}
\includegraphics[width=\columnwidth]{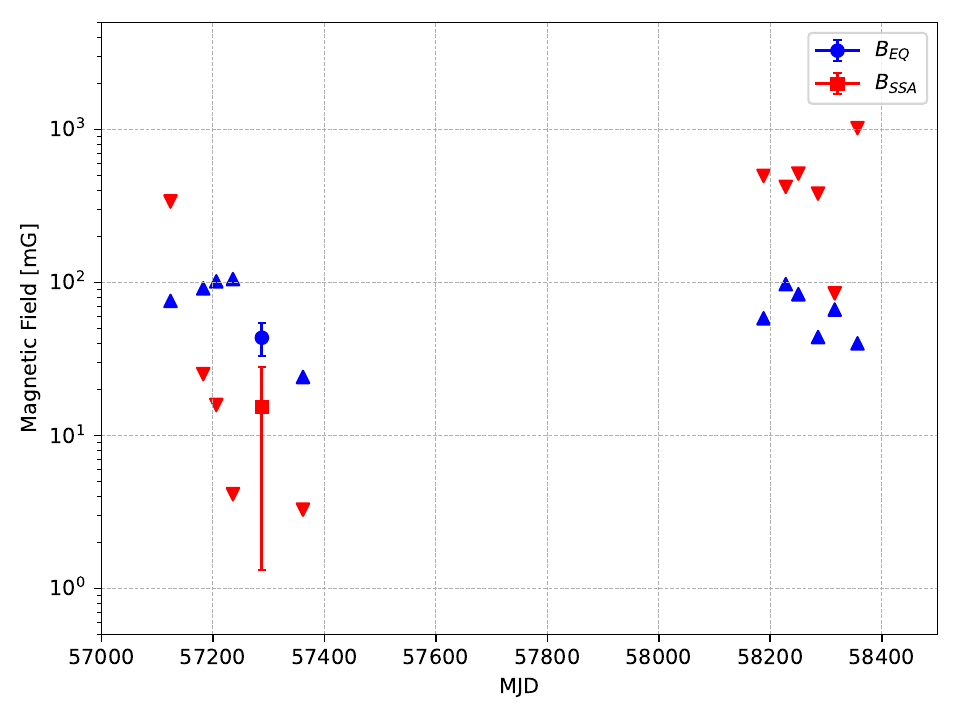}
\caption{Plot of the calculated $B_{\rm SSA}$ and $B_{\rm EQ}$ with time. $B_{\rm SSA}$ is plotted as a red square, and $B_{\rm EQ}$ is plotted as a blue circle. Upper/lower limits are represented by upper/lower triangles respectively. In this plot, we assumed $b(\alpha)=3.8$ and $\eta = 100$.
	\label{fig:BEQBSSA}}
\end{figure}

We estimate the magnetic field strength of AO 0235+164 assuming a synchrotron self-absorption (hereafter SSA) spectrum using the equation \citep{1983ApJ...264..296M}
\begin{equation}
\left[\frac{B_{\rm SSA}}{10^{-2}\textrm{ mG}}\right] 
= b(\alpha)\left[\frac{1 \textrm{ Jy}}{S_{\rm m}}\right]^2\left[\frac{d_{\rm m}}{1 \textrm{ mas}}\right]^4\left[\frac{\nu_{\rm c}}{1\textrm{ GHz}}\right]^5\left(\frac{1+z}{\delta}\right),
\label{eq:bssa}
\end{equation} 
where $b(\alpha)$ is a dimensionless factor that varies as a function of the optically-thin spectral index $\alpha$ \citep[e.g.,][]{1983ApJ...264..296M,2019MNRAS.482.2336P}, $S_{\rm m}$ is the flux density at the turnover frequency $\nu_{\rm c}$, $d_{\rm m}$ is the angular size of the SSA region, $\delta$ is the Doppler factor, and $z$ is the cosmological redshift of the source. The difference in the power of the factor $\left(\frac{\delta}{1+z}\right)$ in Equation \ref{eq:bssa} from that of \citet{1983ApJ...264..296M} is due to considering the SSA region to be the radio core as in \citet{2017ApJ...841..119L}.\\
\indent{}We estimate $d_{\rm m}$ from the model-fit analysis of multi-epoch 43~GHz VLBA data as in Section~\ref{sec:VLBI-JetComp}, but this time focusing on the radio core. For the core, we determine $\theta_{\rm min}$ separately for the major and minor axes of the elliptical Gaussian fitted to the radio core. We consider the core to be unresolved when either of the two are smaller than the corresponding $\theta_{\rm min}$. Following \citet{2017ApJ...834...65A}, the mean size of the fitted elliptical Gaussian is calculated as $\left<\theta_{\rm vlbi}\right> = (2\theta_{\rm major}+\theta_{\rm minor})/3$, where $\theta_{\rm major}$ and $\theta_{\rm minor}$ are the FWHMs of the major and minor axes of the elliptical Gaussian. Then, we have
\begin{equation}
d_{\rm m} = 1.8\left<\theta_{\rm vlbi}\right>\left(\frac{\nu_{\rm c}}{\nu_{\rm obs}}\right)^{-k},
\end{equation}
where $k$ determines the dependence of the core size with frequency. An additional factor of 1.8 is used to convert between the FWHM of a Gaussian function and the total size \citep{1977ApJ...216..244M, 2017A&A...597A..80H}.\\
\indent{} We also estimate the magnetic field based on the assumption of equipartition between the magnetic field and the particle energy densities. Following \citet{2005ApJ...622..797K}, we have
\begin{equation}
\begin{split}
\left[\frac{B_{\rm EQ}}{1\textrm{ mG}}\right] 
=0.123&\left\{\eta^{2}(1+z)^{11}\left[\frac{100\textrm{ Mpc}}{D_{\rm L}}\right]^{2}\left[\frac{\nu_{\rm c}}{5\textrm{ GHz}}\right]\right.\\
&\left.\left[\frac{S_{\rm m}}{0.1\textrm{ Jy}}\right]^{2}\left[\frac{300\textrm{ mas}}{d_{\rm m}}\right]^{6}\delta^{-5}\right\}^{\frac{1}{7}},
\end{split}
\label{eq:Beq}
\end{equation}
where $\eta$ is the ratio of the energy of both the protons and electrons to the energy of the electrons ($\eta = 1$ for a leptonic jet, $\eta = 1836$ for a hadronic jet) and $D_{\rm L}$ is the luminosity distance. The errors $\sigma_{\rm SSA}$ and $\sigma_{\rm EQ}$ for $B_{\rm SSA}$ and $B_{\rm EQ}$ are evaluated with standard error propagation as
\begin{equation}
\left[\frac{\sigma_{\rm SSA}}{B_{\rm SSA}}\right]
= \sqrt{\left[\frac{\sigma_{S_{\rm m}}}{S_{\rm m}}\right]^2+\left[\frac{5\sigma_{\nu_{\rm c}}}{\nu_{\rm c}}\right]^2+\left[\frac{4\sigma_{d_{\rm m}}}{d_{\rm m}}\right]^2+\left[\frac{\sigma_\delta}{\delta}\right]^2}
\label{eq:sig_bssa}
\end{equation}
and
\begin{equation}
\left[\frac{\sigma_{\rm EQ}}{B_{\rm EQ}}\right]
= \sqrt{\left[\frac{2\sigma_{S_{\rm m}}}{7S_{\rm m}}\right]^2+\left[\frac{\sigma_{\nu_{\rm c}}}{7\nu_{\rm c}}\right]^2+\left[\frac{6\sigma_{d_{\rm m}}}{7d_{\rm m}}\right]^2+\left[\frac{5\sigma_\delta}{7\delta}\right]^2},
\label{eq:sig_beq}
\end{equation} 
where $\sigma_{S_{\rm m}}$, $\sigma_{\nu_{\rm c}}$, $\sigma_{d_{\rm m}}$, and $\sigma_\delta$ are the uncertainties of $S_{\rm m}$, $\nu_{\rm c}$, $d_{\rm m}$, and $\delta$.\\
\indent{}To ensure that the spectra from the multiwavelength data of single-dish and compact array observations in Section~\ref{sec:radio_curve_spectrum} represent the radio core, we limit the calculation to VLBI observations where the model-fitted core flux is at least 9 times greater than the sum of all other components. We adopt the value of $\delta=\delta_{\rm var}=28.5\pm8.4$ from Section~\ref{sec:VLBI-JetComp} and $k=0.8\pm0.1$ from \citet{2018MNRAS.475.4994K}. For VLBI observations where the core is unresolved, we use the calculated minimum resolvable size for the major and minor axes to place limits on $B_{\rm SSA}$ and $B_{\rm EQ}$. The results are presented in Figure~\ref{fig:BEQBSSA} and summarized in Table~\ref{tab:Bfields} along with the values of $S_{\rm m}$, $\nu_{\rm c}$, and $d_{\rm m}$. We note that the uncertainties derived for $B_{\rm SSA}$ and $B_{\rm EQ}$ do not account for uncertainties due to $b(\alpha)$, $\eta$, $z$, and $D_{\rm L}$. We find that among the VLBI observations that satisfy the core-dominance criteria, the core was resolved in only one epoch (MJD 57288) with values of $B_{\rm SSA}\approx 
15.3\left(b(\alpha)/3.8\right)\textrm{ mG}$ and $B_{\rm EQ}\approx 43.6\left(\eta/100\right)^{2/7}\textrm{ mG}$. In all other epochs, the core was found to be unresolved, leading to upper limits on $B_{\rm SSA}$ and lower limits on $B_{\rm EQ}$.
\\
\begin{table}
\caption{The magnetic fields of AO 0235+164 assuming $b(\alpha)=3.8$ and $\eta=100$. The columns correspond to (1) the epoch of measurement in MJD, (2) the peak flux density in units of Jy, (3) the turnover frequency in units of GHz, (4) the size of the emission region at the turnover frequency in units of $\mu\textrm{as}$, (5) the SSA magnetic field strength in units of mG, and (6) the equipartition magnetic field strength in units of mG. All uncertainties correspond to 1$\sigma$ confidence regions for the parameters.}
\label{tab:Bfields}
\renewcommand{\arraystretch}{1.5}
\begin{tabular}{cccccc}
\hline
MJD & $S_{\rm m}$ & $\nu_{\rm c}$ & $d_{\rm m}$ & $B_{\rm SSA}$ & $B_{\rm EQ}$\\
$\textrm{[day]}$ & [Jy] & $\textrm{[GHz]}$ & [$\mu$as] & [mG] & [mG]\\
\hline
57124 & $1.85^{+0.14}_{-0.14}$ & $58.4^{+7.22}_{-5.34}$ & $<160$ & $<338$  & $>75.7$\\
57183 & $2.49^{+0.10}_{-0.10}$ & $44.6^{+2.47}_{-2.78}$ & $<135$ & $<25.2$ & $>91.5$\\
57206 & $3.02^{+0.09}_{-0.09}$ & $45.9^{+2.66}_{-2.92}$ & $<128$ & $<15.8$ & $>102$\\
57236 & $3.99^{+0.09}_{-0.09}$ & $38.5^{+2.37}_{-2.65}$ & $<131$ & $<4.13$ & $>105$\\
57288 & $3.43^{+0.07}_{-0.06}$ & $23.1^{+2.63}_{-3.04}$ & $322^{+41}_{-45}$ & $15.3^{+12.6}_{-14.0}$ & $43.6^{+10.6}_{-10.4}$ \\
57362 & $2.30^{+0.18}_{-0.11}$ & $10.3^{+5.38}_{-5.08}$ & $<491$ & $<3.28$ & $>24.1$\\
58188 & $1.44^{+0.15}_{-0.14}$ & $48.9^{+22.3}_{-14.6}$ & $<194$ & $<497$  & $>58.3$\\
58228 & $1.05^{+0.13}_{-0.11}$ & $69.8^{+93.7}_{-39.3}$ & $<103$ & $<422$  & $>97.5$\\
58250 & $1.25^{+0.11}_{-0.11}$ & $65.2^{+8.04}_{-6.86}$ & $<128$ & $<515$  & $>83.8$\\
58286 & $1.43^{+0.09}_{-0.09}$ & $36.9^{+4.11}_{-5.27}$ & $<257$ & $<381$  & $>44.0$\\
58316 & $1.31^{+0.12}_{-0.11}$ & $39.5^{+4.56}_{-6.07}$ & $<157$ & $<84.9$  & $>66.6$\\
58356 & $0.73^{+0.22}_{-0.15}$ & $37.6^{+26.3}_{-21.8}$ & $<230$ & $<1020$ & $>40.0$\\
\hline
\end{tabular}
\end{table}

\section{Discussion}
\subsection{Equipartition of the SSA region}\label{sec:equpartTest}
A number of previous papers \citep{2017ApJ...841..119L, 2018ApJ...859..128A, 2020ApJ...902..104L, 2021A&A...651A..74K, 2022MNRAS.510..815K} estimated both $B_{\rm EQ}$ and $B_{\rm SSA}$ for various sources using similar methods as those used in this paper. They find $B_{\rm EQ}$ values in the range of $0.93\textrm{ mG}$ to $1150\textrm{ mG}$ and $B_{\rm SSA}$ values in the range of $0.01\textrm{ mG}$ to $135\textrm{ mG}$. In all previous cases, the authors found that $B_{\rm EQ}\gg$ $B_{\rm SSA}$, implying that the SSA region in blazars significantly deviates from equipartition. This is in contrast with the results of this paper where the single epoch measurement of the magnetic field strengths implies that the SSA region of AO 0235+164 during the 2015 flare was approximately in an equipartition state.\\
\indent{}We briefly consider the effect of some of the assumptions used when deriving the magnetic field strengths in Section~\ref{sec:bssabeq}. In calculating $B_{\rm SSA}$, we have taken $b(\alpha)=3.8$ (or $\alpha=1$, for $S(\nu)\propto\nu^{-\alpha}$) as a benchmark value. The relatively low cadence of the high-frequency (228, 343~GHz) light curves combined with the highly variable nature of AO 0235+164 makes it difficult to concretely track the variation of $\alpha$. The value of $b(\alpha)$ varies from 3.2 to 3.8 for $\alpha$ of 0.5 to 1.0. The mean spectral index between 228~GHz and 343~GHz during the decay of the 2015 flare is $\sim0.7$, and the change of $B_{\rm SSA}$ due to $b(\alpha)$ is less than the uncertainties due to other observables. The values of $B_{\rm EQ}$ may vary by a factor of $\sim0.27$ (for a fully leptonic jet) to $\sim2.3$ (for a fully hadronic jet) depending on the composition of the jet. Modeling of the broadband spectral energy distribution (SED) of AO 0235+164 suggests a moderately hadronic jet \citep[e.g.,][]{2012ApJ...751..159A} as was considered in our analysis.\\
\indent{}Another method of cross-checking our magnetic field strength estimates is to derive the so-called "time-lag core shift" \citep{2011MNRAS.415.1631K} using the results from Section~\ref{sec:MW_corr} and Section~\ref{sec:VLBI-JetComp}. The time-lag core shift between two frequencies $\nu_{\rm h}$ and $\nu_{\rm l}$ is defined as 
\begin{equation}
\begin{split}
\left[\frac{\Omega_{\rm r\nu}}{1\textrm{ pc GHz}^{1/k_{\rm r}}}\right] = & \frac{4.85\times10^{-3}}{(1+z)^2}\left[\frac{\mu}{1\textrm{ mas yr}^{-1}}\right]\left[\frac{\Delta t}{1\textrm{ yr}}\right]\\
&\times\left[\frac{D_{\rm L}}{1\textrm{ Mpc}}\right]\left(\frac{\nu_{\rm h}^{1/k_{\rm r}}\nu_{\rm l}^{1/k_{\rm r}}}{\nu_{\rm h}^{1/k_{\rm r}}-\nu_{\rm l}^{1/k_{\rm r}}}\right).
\end{split}
\end{equation}
This is a modification of the core shift from the form in \citet{1998A&A...330...79L} by using the proper motion $\mu$ and the observed time lag $\Delta t$ to estimate the angular separation between the radio cores at two different frequencies. The parameter $k_{\rm r}=\left[\left(3+2\alpha\right)m+2n-2\right]/\left(5+2\alpha\right)$, where $m$ and $n$ represent the power-law dependence of the magnetic field ($B(r)\propto r^{-m}$) and the electron number density ($N(r)\propto r^{-n}$) as a function of distance $r$ from the jet base. $\alpha$ is the optically thin spectral index. With an estimate of the core shift, we can define a relation for the distance of the radio core from the jet base for a given frequency $\nu$: $r_{\rm core}(\nu)= \frac{\Omega_{\rm r\nu}}{\sin\theta_{\rm v}}\nu^{-1/k_{\rm r}}$. From our DCF time lags (Section~\ref{sec:MW_corr}), a proper motion of $\mu\approx0.23\textrm{ mas/yr}$ and viewing angle of $\theta_{\rm v}\approx1.42$ degrees (Section~\ref{sec:VLBI-JetComp}), we find that all frequency pairs considered consistently indicate that the 15~GHz core is located at approximately $6\textrm{ pc}$ from the jet base (see Table~\ref{tab:tlagBfields}). The 37~GHz to 343~GHz cores are found to be in the range of $3\textrm{ pc}$ to $
0.4\textrm{ pc}$ from the jet base. \\
\indent{}Assuming equipartition, the magnetic field strength at $1\textrm{ pc}$ from the jet base may be estimated as \citep{2005ApJ...619...73H, 2016A&A...590A..48K}
\begin{equation}
B_{1\textrm{ pc}} = 14\left[\frac{\Omega_{\rm r\nu}^{3k_{\rm r}}(1+z)^2\ln{(\gamma_{\rm max}/\gamma_{\rm min})}}{\delta^2\phi\sin^{3k_{\rm r}-1}{\theta_{\rm v}}}\right]^{0.25}\textrm{ [mG]},
\end{equation}
where $\gamma_{\rm min}$/$\gamma_{\rm max}$ are the minimum/maximum Lorentz factors and $\phi$ is the intrinsic jet half-opening angle. The magnetic field strength at the radio core for a given frequency may also be found from \citep{2016A&A...590A..48K}
\begin{equation}
B_{\rm core}\left(\nu\right) = B_{1\textrm{ pc}}r_{\rm core}^{-1}\left(\nu\right).
\label{eq:Bcore_from_B1pc}
\end{equation}
Adopting the same jet parameters as in Section~\ref{sec:bssabeq} and taking $\phi\approx1\textrm{ deg}$ from \citet{2018MNRAS.475.4994K} and $\ln{(\gamma_{\rm max}/\gamma_{\rm min})}\approx10$, we find that all frequency pairs consistently imply $B_{\rm 1pc}\approx0.2\textrm{ G}$ and $B_{\rm core}\left(15\textrm{ GHz}\right)\approx33\textrm{ mG}$. Evaluating Equation \ref{eq:Bcore_from_B1pc} at the turnover frequency of $\nu_{\rm c}=23\textrm{ GHz}$ for the epoch of MJD 57288, we find $B_{\rm core}\left(23\textrm{ GHz}\right)\approx47\textrm{ mG}$. This is consistent with both $B_{\rm EQ}$ and $B_{\rm SSA}$ found in Section~\ref{sec:bssabeq} for the same epoch.
\begin{table*}
\caption{Time-lag core shifts of AO 0235+164. In all cases, the lower frequency $\nu_{\rm l} = 15\textrm{ GHz}$. The errors of each parameter are calculated from the uncertainty in the time lag $\Delta t$. All uncertainties correspond to 1$\sigma$ confidence regions for the parameters. Systematic errors arising from the jet and cosmological parameters that are common to the various frequency pairs are not included in the calculations. The columns correspond to (1) the paired light-curve frequency in units of GHz, (2) the derived time-lag core shift, (3) the distance of the 15~GHz core from the jet base, (4) the distance of the $\nu_{\rm h}$ core from the jet base, (5) the core shift magnetic field strength at 1~pc from the jet base, (6) the core shift magnetic field strength at the 15~GHz core, and (7) the core shift magnetic field strength at the $\nu_{\rm h}$ core.}
\label{tab:tlagBfields}
\begin{tabular}{ccccccc}
\hline
$\nu_{\rm h}$ & $\Omega_{\rm r\nu}$ & $r_{\rm core}(\nu_{\rm l})$ & $r_{\rm core}(\nu_{\rm h})$ & $B_{1\textrm{pc}}$ & $B_{\rm core}(\nu_{\rm l})$ & $B_{\rm core}(\nu_{\rm h})$\\
$\textrm{[GHz]}$ & [pc GHz$^{0.8}$] & [pc] & [pc] & [mG] & [mG] & [mG]\\
\hline
343 & $1.16\pm0.48$ & $5.37\pm2.22$ & $0.44\pm0.18$ & $181\pm70$\ \  & $33.7\pm0.9$ & $412\pm10.6$\\
228 & $1.88\pm0.78$ & $8.67\pm3.59$ & $0.98\pm0.41$ & $284\pm110$ & $32.7\pm0.8$ & $288\pm7.5$\ \ \\
90  & $1.39\pm0.68$ & $6.45\pm3.14$ & $1.54\pm0.75$ & $215\pm98$\ \  & $33.3\pm1.0$ & $140\pm4.3$\ \ \\
37  & $1.44\pm0.75$ & $6.67\pm3.50$ & $3.24\pm1.70$ & $222\pm109$ & $33.2\pm1.1$ & $68.5\pm2.2$\ \ \ \\
\hline
\end{tabular}
\end{table*}

\begin{figure}
\includegraphics[width=\columnwidth]{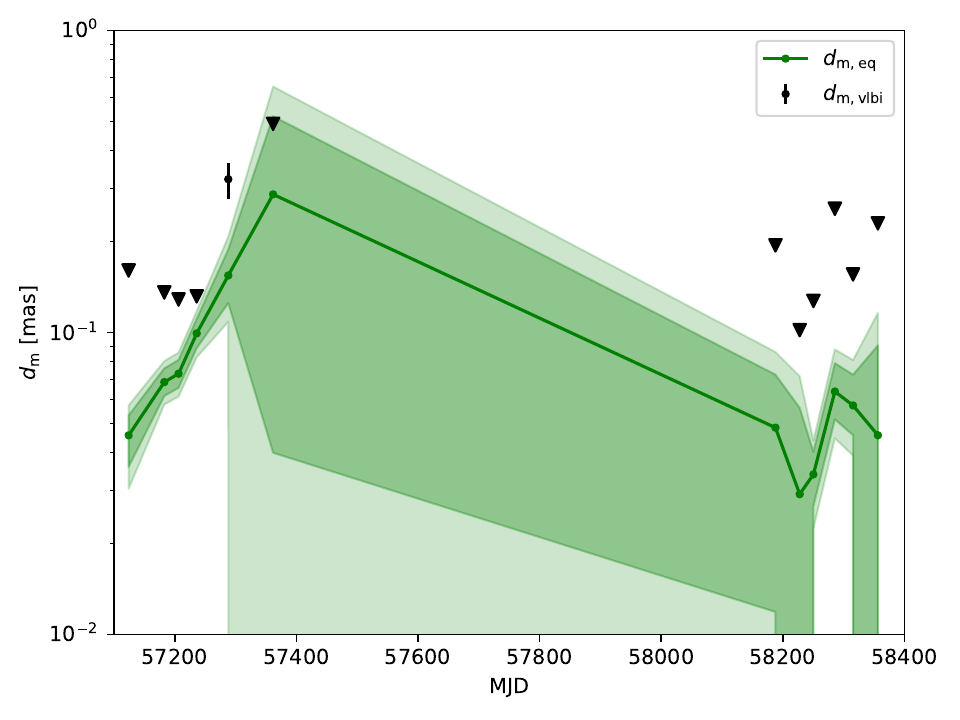}
\caption{A plot of the emission region size derived from VLBI observations ($d_{\rm m, vlbi}$) and the equipartition assumption ($d_{\rm m, eq}$). Upper limits on $d_{\rm m, vlbi}$ are plotted with triangles. The errors on $d_{\rm m, vlbi}$ represent 1$\sigma$ confidence regions. The 90-percent and 99-percent confidence regions of $d_{\rm m,eq}$ are plotted as filled regions of darker and lighter green respectively.
	\label{fig:equipartsize}}
\end{figure}

We may define an ``equipartition size" by equating $B_{\rm SSA}$ and $B_{\rm EQ}$. From Equations \ref{eq:bssa} and \ref{eq:Beq}, we find 
\begin{equation}
\begin{split}
\left[\frac{d_{\rm m, eq}}{1\textrm{ mas}}\right]\approx&1.94\left\{\frac{\eta^2\delta^2\left(1+z\right)^{4}}{b(\alpha)^7}\left[\frac{100\textrm{ Mpc}}{D_{\rm L}}\right]^{2}\right\}^{\frac{1}{34}}\\
&\times\left[\frac{1\textrm{ GHz}}{\nu_{\rm c}}\right]\left[\frac{S_{\rm m}}{1 \textrm{ Jy}}\right]^\frac{8}{17}.
\end{split}
\end{equation}
The uncertainties are evaluated using error propagation,
\begin{equation}
\left[\frac{\sigma_{d_{\rm m, eq}}}{d_{\rm m, eq}}\right]=\sqrt{\left[\frac{\sigma_{\nu_{\rm c}}}{\nu_{\rm c}}\right]^2+\left[\frac{8\sigma_{S_{\rm m}}}{17S_{\rm m}}\right]^2+\left[\frac{\sigma_{\delta}}{17\delta}\right]^2}.
\end{equation}
We evaluate the above equations for our core-dominated epochs and compare them to the VLBI-derived SSA region sizes $d_{\rm m,vlbi}$. The results are presented in Figure~\ref{fig:equipartsize}. We find that $d_{\rm m, eq}$ varies between $0.03\textrm{ mas}$ and $0.3\textrm{ mas}$ during our period of interest. We also see that the observed upper limits on $d_{\rm m,vlbi}$ are consistent with the values of $d_{\rm m, eq}$.

\subsection{Shock-in-jet interpretation of the 2015 flare}\label{sec:shock-in-jet}
As shown in Section~\ref{sec:radio_curve_spectrum}, the evolution of $S_{\rm m}$ and $\nu_{\rm c}$ may be decomposed into different stages, starting with a factor of $\sim$2 rise of $S_{\rm m}$ at roughly constant $\nu_{\rm c}$, followed by a $\sim$50\% reduction of $\nu_{\rm c}$ with less variation of $S_{\rm m}$ (remaining within $\sim$10\% of 4~Jy), then finally a decrease of $S_{\rm m}$ by $\sim$50\% at an approximately constant $\nu_{\rm c}$. This seems to imply that the flare may be interpreted according to the shock-in-jet model \citep{1985ApJ...298..114M} with the three stages each corresponding to the Compton, synchrotron, and adiabatic stage respectively. We find similar trends for the 2018 flare; however, the larger uncertainties complicate the identification of the three stages. Within the shock-in-jet scenario, we may interpret the newly ejected jet component J5 (and to a lesser certainty, J6) found in the 43~GHz VLBA maps (Section~\ref{sec:VLBI-JetComp}) to be the propagating relativistic shock that produced the 2015 flare. It should be noted that the exact multiwavelength characteristics of radio flares predicted by the shock-in-jet scenario depends greatly on the physical parameters of the jet, such as the energy distribution of the synchrotron-emitting electrons, and the change of the magnetic field strength, particle density, and Doppler factor with distance from the jet base \citep[e.g.,][]{2011A&A...531A..95F, 2015A&A...580A..94F}. It may be that the Compton stage extends to MJD~57248, with $S_{\rm m}$ peaking at 4.4~Jy and a marginal decrease of $\nu_{\rm c}$ to 35~GHz. The synchrotron stage then may result in a decrease of $S_{\rm m}$ to 3.7~Jy and a decrease of $\nu_{\rm c}$ to 25~GHz, followed by the adiabatic stage starting at MJD 57270. Detailed modeling of the evolution of $S_{\rm m}$ and $\nu_{\rm c}$, or of the multiwavelength radio light curves themselves, may in turn allow for constraints on the jet parameters of AO 0235+164.\\
\indent{}The distance $\Delta d$ between the radio and $\gamma$-ray emission regions may be estimated as \citep{2010ApJ...722L...7P}
\begin{equation}
\Delta d = \frac{\beta_{\rm app}c\tau_{\rm DCF}}{(1+z)\sin\theta_{\rm v}}.
\end{equation}
Using $\tau_{\rm DCF}$ found from the $\gamma$-ray and the 15~GHz data, we find that the $\gamma$-ray emission region is located $\Delta d \approx 9.9\pm6.1\textrm{ pc}$ upstream of the 15~GHz radio core. This implies that the $\gamma$-rays were produced near the jet base. \citet{2014ApJ...789..161N} use multiple constraints to infer that the 2008 $\gamma$-ray flare was produced at $\sim1$~pc from the central engine while the 43~GHz VLBI core is at $\sim6.7$~pc. This is consistent with what we find for the 2015 flare where we find the $\gamma$-ray production region to be $\sim6.5$~pc upstream of the 37~GHz core. The inferred deprojected distance between the 37~GHz core and the 43~GHz core is $<0.5$~pc, which is within the measurement uncertainties. From these distances and the radio-$\gamma$-ray correlation, we can infer that the 2015 flare may have been caused by a relativistic shock propagating along the jet. Particle acceleration by this shock near the jet base may have produced the leading $\gamma$-ray flare followed by the subsequent radio flares (with lower frequencies peaking later) as the shock propagates further downstream and interacts with the radio core. It should be noted that the uncertainty in the location of the $\gamma$-ray flaring region is large, primarily due to the large uncertainty in the $\gamma$-to-radio time lag. Future high-cadence monitoring, as well as $\gamma$-ray observations at increased sensitivity (enabling detection within smaller time bins), may help to better constrain the time delay between $\gamma$-ray and radio light curves. In turn, better constraints in the time delay will enable better constraints in the $\gamma$-ray emission region.\\
\indent{}An alternative mechanism that may produce strong $\gamma$-ray flares followed by delayed radio flares is magnetic reconnection \citep[e.g.,][]{2016MNRAS.462.3325P}. Emission models based on particle-in-cell simulations of magnetic reconnection are able to reproduce the multiwavelength SEDs of both low-frequency and high-frequency peaked blazars \citep{2016MNRAS.462.3325P}, and of both BL Lacs and flat spectrum radio quasars \citep{2019MNRAS.482...65C}. $\gamma$-ray flares from magnetic reconnection are expected to vary on timescales of several hours to days, depending on the size of the produced plasmoids. In this paper, we investigate the $\gamma$-ray variability of AO 0235+164 with 7-day time bins, which lacks the temporal resolution to study flares on such timescales. The radio synchrotron emission from plasmoids created by magnetic reconnection events is expected to be strongly attenuated by SSA and unlikely to be observed directly \citep{2016MNRAS.462.3325P}. The radio emission may be observable once the particles escape the reconnection layer \citep{2019MNRAS.482...65C}. A detailed study of the temporal and spectral evolution of radio flares following magnetic reconnection may further shed light on whether such processes are able to explain the multiwavelength flares of AO 0235+164.

\subsection{Core dominance of the radio flux}\label{sec:SD-Core_Dominance}
\begin{figure}
\includegraphics[width=\columnwidth]{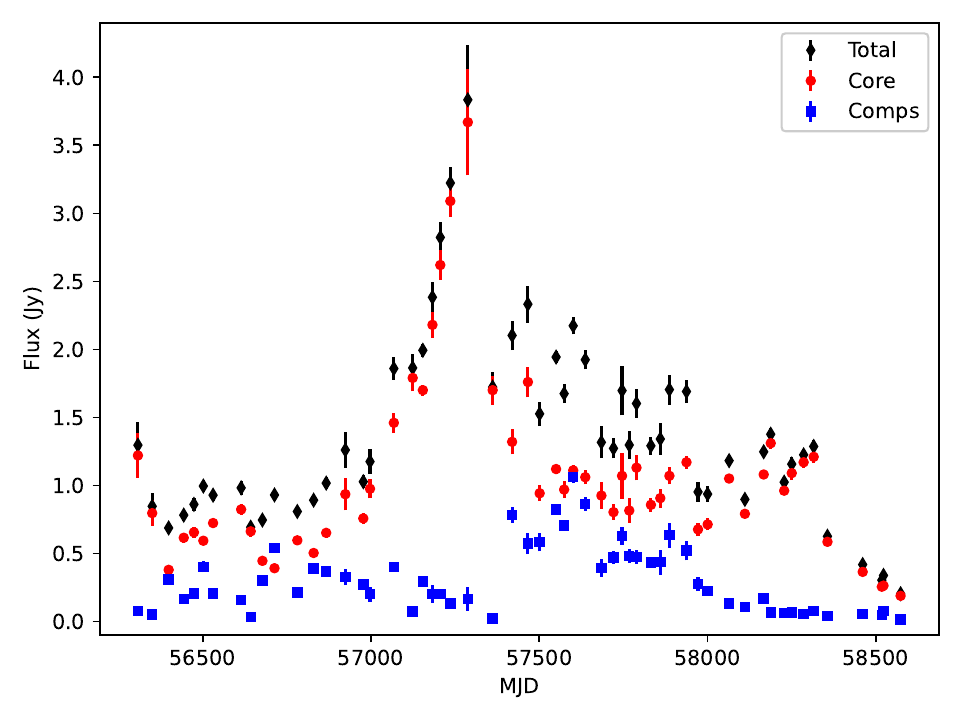}
\caption{Plot of the 7mm VLBI light curve. The core component is plotted as red dots while the sum of the remaining components is plotted as blue squares. The total VLBI flux is plotted as black diamonds.
	\label{fig:VLBICoreCompLC}}
\end{figure}

AO 0235+164 is compact with minimal apparent structure extending past a couple mas from the center. VLBI observations at cm wavelengths find the flux of AO 0235+164 to be core dominated \citep[e.g.,][]{2016AJ....152...12L, 2018MNRAS.475.4994K}. On the other hand, multi-epoch mm-VLBI observations are able to resolve the sub-mas jet of this source and find that for certain epochs, the jet components account for a considerable fraction of the total flux density \citep[e.g.,][]{2006ApJ...640..196P, 2017ApJ...846...98J}. We find this to be the case during our time period of interest. As shown in Figure~\ref{fig:VLBICoreCompLC}, the model fits to the 43~GHz VLBA data suggest that the combined flux for the extended structures may account for up to half of the total flux in some epochs. However, we note that from the time period of the 2015 flare until the jet component is resolved from the core in 2016, the total contribution of any external components is only a few percent. Therefore, we consider the multiwavelength total flux measurements to be good representations of the radio core flux during this period.

\section{Conclusion}
In this paper, we investigated the multiwavelength flaring activity of AO 0235+164 during its recent active period from 2013 to 2019. From a DCF time-lag analysis, we find that the radio and $\gamma$-ray light curves are correlated with a significance of $2\sigma$ or greater, which has been previously demonstrated for this source \citep[e.g.,][]{2011ApJ...735L..10A,2014MNRAS.445..428M}. We also find that variability at lower frequencies is delayed with respect to variability at higher frequencies, implying that the radio variability of AO 0235+164 is affected by the opacity of the radio core (which is responsible for most of the radio flux variability). This is consistent with the results of \citet{2014MNRAS.441.1899F} who performed a cross-correlation analysis of a sample of blazars (including AO 0235+164) that are bright in the \textit{Fermi}-LAT band. They also found a trend of increasing lags toward lower frequencies with source frame time delays on the order of days$-$tens of days. We identify a newly ejected superluminal $(\beta_{\rm app}\approx12\textrm{ c})$ jet component associated with the 2015 flare. Tracking the time evolution of this jet component, we find $\delta_{\rm v}=29$, $\Gamma=17$, and $\theta_{\rm v}=1.4\textrm{ degrees}$. We evaluated the SSA spectrum from quasi-simultaneous radio spectra constructed with GPR, finding the turnover frequency $\nu_{\rm c}$ to vary between 10~GHz and 70~GHz and the peak flux density $S_{\rm m}$ to vary between 0.7~Jy and 4~Jy. The time-evolution trends in the $S_{\rm m}$-$\nu_{\rm c}$ parameter space could be consistent with a shock-in-jet origin of the flaring activity. We evaluated $B_{\rm SSA}$ and $B_{\rm EQ}$, finding upper limits on $B_{\rm SSA}$ in the range of $4\textrm{ mG}$ to $1020\textrm{ mG}$ and lower limits on $B_{\rm EQ}$ in the range of $24\textrm{ mG}$ to $105\textrm{ mG}$. A single measurement of $B_{\rm SSA}=15.3^{+12.6}_{-14.0}\textrm{ mG}$ and $B_{\rm EQ}=43.6^{+10.6}_{-10.4}\textrm{ mG}$ was obtained close to the peak of the 2015 radio flare when the radio core was resolved in 43~GHz VLBI data. From additional investigation of the time-lag core shifts and the equipartition size, we find that the SSA region of AO 0235+164 is consistent with the equipartition assumption. We infer that the SSA region was approximately $1\textrm{ pc}$ - $10\textrm{ pc}$ downstream of the jet base during the 2015 flare while the $\gamma$-rays were produced close to the jet base.\\

\section*{Acknowledgements}
We thank the anonymous reviewer for valuable comments and suggestions that helped to improve the paper. The manuscript was improved by the helpful comments of Filippo D'Ammando, Tonia Venters, Dave Thompson and Deirdre Horan.
This research has made use of data from the OVRO 40-m monitoring program (Richards, J. L. et al. 2011, ApJS, 194, 29), supported by private funding from the California Insitute of Technology and the Max Planck Institute for Radio Astronomy, and by NASA grants NNX08AW31G, NNX11A043G, and NNX14AQ89G and NSF grants AST-0808050 and AST- 1109911. This publication makes use of data obtained at Mets\"ahovi Radio Observatory, operated by Aalto University in Finland. JYK was supported for this research by the National Research Foundation of Korea (NRF) funded by the Korean government (Ministry of Science and ICT; grant no. 2022R1C1C1005255). I.A. acknowledges financial support from the Spanish "Ministerio de Ciencia e Innovaci\'on” (MCINN) through the “Center of Excellence Severo Ochoa” award for the Instituto de Astrof\'isica de Andaluc\'ia-CSIC (SEV-2017-0709). Acquisition and reduction of the MAPCAT data was supported in part by MICINN through grants AYA2016-80889-P and PID2019-107847RB-C44. The POLAMI observations were carried out at the IRAM 30m Telescope. IRAM is supported by INSU/CNRS (France), MPG (Germany) and IGN (Spain ). CC acknowledges support from the European Research Council (ERC) under the HORIZON ERC Grants 2021 program under grant agreement No. 101040021. This paper makes use of the following ALMA data: ADS/JAO.ALMA\#2011.0.00001.CAL. ALMA is a partnership of ESO (representing its member states), NSF (USA) and NINS (Japan), together with NRC (Canada), MOST and ASIAA (Taiwan), and KASI (Republic of Korea), in cooperation with the Republic of Chile. The Joint ALMA Observatory is operated by ESO, AUI/NRAO and NAOJ. The Submillimeter Array is a joint project between the Smithsonian Astrophysical Observatory and the Academia Sinica Institute of Astronomy and Astrophysics and is funded by the Smithsonian Institution and the Academia Sinica. We recognize that Maunakea is a culturally important site for the indigenous Hawaiian people; we are privileged to study the cosmos from its summit. This study makes use of VLBA data from the VLBA-BU Blazar Monitoring Program (BEAM-ME and VLBA-BU-BLAZAR;
http://www.bu.edu/blazars/BEAM-ME.html), funded by NASA through the Fermi Guest Investigator Program. The VLBA is an instrument of the National Radio Astronomy Observatory. The National Radio Astronomy Observatory is a facility of the National Science Foundation operated by Associated Universities, Inc.\\
The \textit{Fermi} LAT Collaboration acknowledges generous ongoing support
from a number of agencies and institutes that have supported both the
development and the operation of the LAT as well as scientific data analysis.
These include the National Aeronautics and Space Administration and the
Department of Energy in the United States, the Commissariat \`a l'Energie Atomique
and the Centre National de la Recherche Scientifique / Institut National de Physique
Nucl\'eaire et de Physique des Particules in France, the Agenzia Spaziale Italiana
and the Istituto Nazionale di Fisica Nucleare in Italy, the Ministry of Education,
Culture, Sports, Science and Technology (MEXT), High Energy Accelerator Research
Organization (KEK) and Japan Aerospace Exploration Agency (JAXA) in Japan, and
the K.~A.~Wallenberg Foundation, the Swedish Research Council and the
Swedish National Space Board in Sweden.\\
Additional support for science analysis during the operations phase is gratefully
acknowledged from the Istituto Nazionale di Astrofisica in Italy and the Centre
National d'\'Etudes Spatiales in France. This work performed in part under DOE
Contract DE-AC02-76SF00515.\\
This research made use of Astropy,\footnote{http://www.astropy.org} a community-developed core Python package for Astronomy \citep{2013A&A...558A..33A, 2018AJ....156..123A}. This work was supported by the National Research Foundation of Korea (NRF) grant funded by the Korea government (MIST) (2020R1A2C2009003). 
\section*{Data Availability}

The data underlying this article will be shared on reasonable request to the corresponding author. The \textit{Fermi}-LAT photon data are available through query at the LAT data server (https://fermi.gsfc.nasa.gov/cgi-bin/ssc/LAT/LATDataQuery.cgi). The 15~GHz OVRO data may be available on request to the OVRO 40~m collaboration. For questions regarding the availability of the 37~GHz Metsähovi data, please contact agn-metsahovi at aalto.fi. The ALMA data are available at the ALMA Calibrator Source Catalogue, found at https://almascience.eso.org/sc/. The SMA data are available at http://sma1.sma.hawaii.edu/callist/callist.html. Questions regarding the data and data policies should be addressed to Mark Gurwell (mgurwell at cfa.harvard.edu). The 43 GHz VLBA data from the VLBA-BU-BLAZAR program are available at https://www.bu.edu/blazars/VLBAproject.html.



\bibliographystyle{mnras}
\bibliography{WheeYeon_0235_Draft} 








\bsp	
\label{lastpage}
\end{document}